\documentclass[11pt]{scrartcl}  
\usepackage[left=1in, right=1in, top=1in, bottom=1in]{geometry}
\usepackage{mathptmx} 
\usepackage{multirow,graphicx,physics,hyperref,amsmath,amssymb,graphicx,mathalfa,mathtools,colortbl,hhline,bbold}
\numberwithin{equation}{section}
\hypersetup{colorlinks=true, citecolor=blue}
\usepackage{array}
\usepackage[most]{tcolorbox}
\newtcolorbox{reviewbox}[1]{
    colback=blue!5, colframe=blue!60!black,
    title={#1}, center title, fonttitle=\bfseries, breakable}
\usepackage[framemethod=TikZ]{mdframed}
\mdfdefinestyle{sid}{%
    linecolor=black,
    outerlinewidth=1.0pt,
    roundcorner=7pt,
    innerrightmargin=15pt,
    innerleftmargin=15pt,
    backgroundcolor=gray!30
		}	
\definecolor{Gray}{gray}{0.8}
\definecolor{MyBlue}{rgb}{0.0,0.0,0.9}
\definecolor{MyRed}{rgb}{0.0,0.9,0.0}
\colorlet{Bluee}{MyBlue!6}
\colorlet{Redd}{MyRed!1}
\DeclareMathAlphabet{\mathcal}{OMS}{cmsy}{m}{n} 

\usepackage{xcolor}
\colorlet{sectioncolor}{blue!20}
\colorlet{subsectioncolor}{orange!70}
\colorlet{subsubsectioncolor}{green!40}
\makeatletter
\renewcommand\sectionlinesformat[4]{%
  \colorbox{#1color}{%
    \parbox[t]{\dimexpr\textwidth-2\fboxsep\relax}{%
      \raggedsection\color{black}\@hangfrom{#3}{#4}%
}}}
\makeatother
\usepackage{blindtext}
\title{\textbf{Semiclassical Asymptotics of Entanglement in Torus-Link States}}
\date{}
\author{\textbf{\emph{Vikash Ganwariya}, \emph{Simran Sain}}, \textbf{\emph{Siddharth Dwivedi}}\thanks{Email: 2025phdphy003@curaj.ac.in, 2025phdphy002@curaj.ac.in, siddharth.dwivedi@curaj.ac.in}\\\\ Department of Physics, School of Physical Sciences, \\ Central University of Rajasthan, Ajmer, 305817, India}
\begin{document}
\maketitle
\begin{abstract}
We investigate the entanglement entropy of topological quantum states associated with $T_{d,\,dn}$ torus links in (2+1)-dimensional SU(2) Chern--Simons theory at level $k$. These links consist of $d$ components with pairwise linking number $n$, and the corresponding link-complement states provide a natural setting to study multipartite entanglement in topological quantum field theories. We analyze the semiclassical limit $k\to\infty$ of the entanglement entropy by deriving the asymptotic behavior of the moments of the reduced density matrix. To achieve this, we decompose the integrable-representation space into left-edge, right-edge, and bulk regions and employ the stationary-phase method to evaluate their respective contributions. We find that the leading asymptotic behavior is governed entirely by the two edge regions, while the bulk contribution is asymptotically suppressed. The resulting expression for the limiting entanglement entropy is given in terms of Hurwitz zeta functions and exhibits a dependence on the residue class of $k$ modulo $n$. We further investigate the dependence of the limiting entanglement entropy on the winding parameter $n$ and the number of link components $d$. In the large-$d$ limit, the entropy reduces to a binary entropy determined by two normalized weights associated with the left and right edges. Finally, we show that the two iterated limits, $k\to\infty$ and $d\to\infty$, commute, reproducing the result obtained previously by taking the large-party limit first.
\end{abstract}
\newpage
\hypersetup{linkcolor=blue}
\tableofcontents

\hypersetup{linkcolor=red}
\listoffigures
\hypersetup{linkcolor=brown}
\listoftables
\newpage
\hypersetup{linkcolor=red}
\section{Introduction} \label{sec1}

Quantum entanglement is a fundamental feature of quantum systems and plays a central role in quantum information theory, quantum computation, and the characterization of strongly correlated quantum systems. Despite its importance, studying entanglement in generic quantum field theories is often technically challenging, owing to the infinitely many degrees of freedom and the nontrivial dependence on the geometry of the underlying spacetime. In this regard, topological quantum field theories (TQFTs) provide a particularly useful setting for studying quantum entanglement. Since physical observables and correlation functions in a TQFT are independent of the spacetime metric, such theories isolate the topological aspects of quantum correlations and provide an analytically tractable framework for investigating how entanglement encodes global and topological information.

Chern--Simons theory in (2+1) dimensions \cite{Witten:1988hf} provides a prominent realization of a TQFT and establishes a remarkable connection between quantum field theory, knot and link invariants, and the topology of three-manifolds. In particular, the path integral of Chern--Simons theory on a three-manifold with multiple boundary components naturally prepares a multipartite quantum state in the tensor product of the Hilbert spaces associated with these boundaries. Knot and link complements therefore furnish a rich class of geometrically and topologically nontrivial quantum states whose entanglement properties can be studied systematically. Investigating quantities such as the Rényi and von Neumann entropies of these states can provide insight into how the topology of knots, links, and three-manifolds is encoded in quantum correlations, while also offering a controlled setting for exploring more general connections between topology and quantum entanglement. Over the past decade, several works have investigated different aspects of topological entanglement measures for such multi-boundary states. A non-exhaustive list of relevant studies includes \cite{Balasubramanian:2016sro,Dwivedi:2017rnj,Balasubramanian:2018por,Salton:2016qpp,Hung:2018rhg,Zhou:2019ezk,Dwivedi:2020jyx,Dwivedi:2020rlo,Dwivedi:2021dix,Dwivedi:2024gzg,Saini2025,Cummings:2025zfe,Yuan:2025dgx,Akella:2026xza,Sain:2026ryl}. 

An important feature of quantum states associated with link complements is that their probability amplitudes are directly related to quantum invariants of the corresponding links \cite{Balasubramanian:2016sro}. More precisely, these amplitudes are given by expectation values of Wilson line operators supported on the links, which can be expressed in terms of the associated quantum link invariants. For instance, in SU(2) Chern--Simons theory, these amplitudes are related to colored Jones polynomials evaluated at a root of unity, $q=\exp\left(\frac{2\pi i}{k+2}\right)$, where $k$ denotes the Chern--Simons level. Consequently, the entanglement measures of the resulting quantum states are determined by suitable combinations of these quantum invariants, thereby establishing a direct connection between quantum entanglement and the topological properties of links.

An important direction in this line of research is to investigate the asymptotic behavior of entanglement measures in the semiclassical regime, corresponding to the large-level limit $k \to \infty$. This limit is particularly interesting because it provides a bridge between quantum invariants and the underlying classical and geometric structures of the theory. Moreover, the asymptotic behavior of quantum invariants often reveals unexpected connections with geometry and number theory, involving quantities associated with moduli spaces, special values of zeta functions, and related arithmetic structures. Studying the large-$k$ behavior of entanglement measures therefore offers a natural way to explore whether such geometric and arithmetic information is also encoded in the entanglement structure of link-complement states.

Among the various classes of link complements, in this work we focus on torus links, and in particular on the family $T_{d,\,dn}$. This family provides a useful setting for our analysis because the corresponding link-state coefficients can be expressed in terms of modular matrices, allowing the entanglement structure of the associated states to be investigated analytically. The primary goal of this work is to determine the semiclassical (large-$k$) limit of the entanglement entropy for the $T_{d,\,dn}$ family in SU(2) Chern--Simons theory. Throughout this work, we refer to this large-$k$ limit as the `\emph{limiting entanglement entropy}' and denote it by $\mathrm{EE}_\infty$. 

Earlier works investigating the entanglement structure of torus-link states include  \cite{Dwivedi:2020rlo,Dwivedi:2024gzg,Saini2025,Caputa:2024qkk}. In particular, Refs.~\cite{Dwivedi:2020rlo,Caputa:2024qkk} have made progress towards analyzing the large-$k$ limits of entanglement measures for $T_{d,\,dn}$ link complements. For example, Ref.~\cite{Dwivedi:2020rlo} obtained the large-$k$ limits of the entropies for $n=1,2,3,4,6$. This analysis relied on the observation that the traces of the $m^{\text{th}}$ powers (or moments) of the unnormalized reduced density matrix form sequences of positive integers as the level $k$ is varied. Exploiting this property, the moments were expressed as polynomial functions of $k$ for each Rényi index $m$. By identifying the dependence of the leading-order polynomial coefficients on $m$, the large-$k$ limit of the $m^{\text{th}}$ Rényi entropy was obtained, and subsequently taking the limit $m\to1$ yielded the limiting entanglement entropy. This approach, however, relies crucially on identifying the functional dependence of the polynomial coefficients on the Rényi index $m$. Such a dependence could not be established for $n=5$, or more generally for larger values of $n$, thereby limiting the applicability of this method.

In the present work, we employ Poisson summation and the method of stationary phase to derive the large-$k$ asymptotics of the eigenvalues of the unnormalized reduced density matrix for arbitrary $n$. A key finding is that the leading asymptotic behavior of the density-matrix moments is governed by contributions from the two edges of the integrable-representation space. More specifically, only eigenvalues associated with the left edge $R/k\to 0$ and the right edge $R/k\to 1$ contribute at leading order, whereas the contribution from the bulk $0<R/k<1$ is asymptotically suppressed. This edge-dominance mechanism enables us to derive the large-$k$ limit of the entanglement entropy for the entire $T_{d,\,dn}$ family, for arbitrary $n$. Remarkably, the resulting expressions for the limiting entropies involve finite sums of Hurwitz zeta functions, with coefficients that retain information about the residue class of $k$ modulo $n$.

The paper is organized as follows. In Section~\ref{sec2}, we introduce the theoretical framework of SU(2)$_k$ Chern--Simons theory, construct the quantum states associated with link complements, and specialize to the $T_{d,\,dn}$ torus-link family. In Section~\ref{sec3}, we derive the large-$k$ asymptotics of the density-matrix moments by analyzing the contributions from the left-edge, right-edge, and bulk regions of the integrable-representation space. In Section~\ref{sec4}, we use these asymptotic results to obtain the large-$k$ limits of the Rényi and the entanglement entropies. In Section~\ref{sec5}, we present numerical and analytical results illustrating the dependence of the limiting entanglement entropy on the winding parameter $n$, the number of link components $d$, and the residue class of $k$ modulo $n$. We also examine the large-$d$ behavior and its relation to the semiclassical limit. Finally, we summarize our results and discuss possible directions for future work in Section~\ref{sec6}.

\section{Set-up \label{sec2}}
\subsection{Chern-Simons theory and link complement states}
The Chern-Simons action for a gauge group $G$ at level $k$ on a three-manifold $M$ is given by
\begin{equation}
S(A)=\frac{k}{4\pi}\int_M \operatorname{Tr}\left(A\wedge dA+\frac{2}{3}A\wedge A\wedge A\right),
\end{equation}
where $A=A_\mu dx^\mu$ is the gauge connection on a principal $G$-bundle over $M$. The fundamental gauge-invariant observables of Chern--Simons theory are Wilson loop operators. For an oriented knot $\mathcal K\subset M$, carrying a representation $R$ of $G$, the Wilson loop is defined by:
\begin{equation}
W_R(\mathcal K)=\operatorname{Tr}_R\,P\exp\left(i\oint_{\mathcal K}A\right).
\end{equation}
In the absence of Wilson loop insertions, the partition function of the manifold is
\begin{equation}
Z(M)=\int e^{iS(A)}\,dA ~,
\end{equation}
which is a topological invariant of $M$. If $M$ contains a link
$\mathcal L=\mathcal K_1\sqcup\mathcal K_2\sqcup\cdots\sqcup\mathcal K_d$,
whose components are colored by representations $R_1,R_2,\ldots,R_d$, the corresponding path integral becomes
\begin{equation}
Z(M;\mathcal L)=\int e^{iS(A)}
\prod_{i=1}^{d}W_{R_i}(\mathcal K_i)\,dA.
\end{equation}
For a closed three-manifold, this partition function computes the expectation value of the Wilson loops. By Witten's seminal result \cite{Witten:1988hf}, these expectation values coincide with the quantum link invariants.

When the manifold has a boundary $\Sigma$, the path integral is performed with the boundary condition $A|_\Sigma=Q$. The resulting partition function is interpreted as a quantum state,
\begin{equation}
|\Psi\rangle\equiv Z_Q(M;\mathcal L) = \int_{A|_\Sigma=Q}
e^{iS(A)}\,
\prod_{i=1}^{d}
W_{R_i}(\mathcal K_i)\,
dA \,
\end{equation}
which belongs to the Hilbert space $\mathcal H_\Sigma$ associated with the boundary surface. Throughout this work, we take the boundary to be a torus, $\Sigma=T^2$. The Hilbert space $\mathcal H_{T^2}$ admits a natural basis obtained by performing the Chern-Simons path integral on a solid torus with a Wilson line in representation $R$ inserted along its non-contractible cycle. The allowed labels $R$ are the integrable representations of the affine Lie algebra associated with $G$. For the SU(2) group, these are the representations whose Dynkin label $a$ satisfies the condition $0\leq a \leq k$. Consequently, $\mathcal H_{T^2}$ is finite-dimensional, with a basis
\begin{equation}
\text{basis}(\mathcal{H}_{T^2}) = \{\ket{e_a} : 0\leq  a \leq  k \} ~.
\label{basisHT2SU2}
\end{equation}
Each basis state $\ket{e_a}$ is obtained by the path integral on a solid torus carrying a Wilson line in representation $a$ along its non-contractible cycle.

A manifold $M$ with $d$ disjoint torus boundaries can be interpreted as a $d$-party quantum state which belongs to the tensor product Hilbert space of $d$ copies of $\mathcal{H}_{T^2}$ and can be written as: 
\begin{equation}
\ket{\Psi} = \sum_{a_1,\,\ldots,\,a_d=0}^k C_{a_1,a_2,\ldots,a_d} \, \ket{e_{a_1},e_{a_2},\ldots,e_{a_d}} ~,
\label{Genstate}
\end{equation}
where $\ket{e_{a_1},e_{a_2},\ldots,e_{a_d}}$ is shorthand for $\ket{e_{a_1}} \otimes \ket{e_{a_2}} \otimes \ldots \otimes \ket{e_{a_d}}$. The complex coefficients $C_{a_1,\,a_2,\ldots,\,a_d}$ encode the topological information of the three-manifold $M$ and determine the entanglement properties of the associated quantum state.

When $M = S^3 \backslash \mathcal{L}$ is a link complement, where $\mathcal{L}$ is a $d$-component link, then the coefficients $C_{a_1,\,a_2,\ldots,\,a_d}$ are obtained using the gluing technique prescribed in \cite{Balasubramanian:2016sro} and are identified with the Chern-Simons partition function of $S^3$ in presence of the link $\mathcal{L}$. We call such a state the link complement state and denote it as $\ket{\mathcal{L}}$ given by:
\begin{equation}
\ket{\mathcal{L}} = \sum_{a_1,\,\ldots,\,a_d=0}^k \, Z(S^3;\, \mathcal{L}[a_1,a_2,\ldots,a_d]) \, \ket{e_{a_1},e_{a_2},\ldots,e_{a_d}} ~,
\label{LinkComplstate}
\end{equation}
where $Z(S^3; \, \mathcal{L}[a_1,a_2,\ldots,a_d])$ denotes the $S^3$ partition function in the presence of link $\mathcal{L}$ whose components are colored by $a_1,a_2,\ldots,a_d$. For SU(2) Chern-Simons theory, these partition functions are proportional to the colored Jones polynomials of $\mathcal{L}$ evaluated at the root of unity $q=\exp\left(\frac{2\pi i}{k+2}\right)$. Thus, for SU(2) Chern-Simons theory, the amplitudes of the link complement state are completely determined by the colored Jones polynomials of the corresponding link.
\subsection{Entanglement structure of link complement states}
To obtain the bipartite entanglement structure of the link complement state \eqref{LinkComplstate}, we bipartition the total system as $(A|B) = (p|d-p)$, where Alice's subsystem has $p$ parties, and Bob's system has $(d-p)$ parties. The corresponding bipartitioning of the total Hilbert space will be:
\begin{equation}
\mathcal{H} =
\underbrace{\mathcal{H}_{T^2}\otimes\mathcal{H}_{T^2}\otimes\cdots\otimes
\mathcal{H}_{T^2}}_{p} \,
\otimes \,
\underbrace{\mathcal{H}_{T^2}\otimes\mathcal{H}_{T^2}\otimes\cdots\otimes
\mathcal{H}_{T^2}}_{d-p}
\equiv
\mathcal{H}_{A}\otimes\mathcal{H}_{B}~.
\end{equation}
The total density matrix for the link state will be:
\begin{equation}
\rho_{\text{total}} =
\frac{\ket{\mathcal{L}}\bra{\mathcal{L}}}
{\bra{\mathcal{L}}\ket{\mathcal{L}}} ~.
\end{equation}
The bipartite measures are obtained by taking the partial trace over any one of the subsystems. Taking the partial trace of $\rho_{\text{total}}$ with respect to $\mathcal{H}_B$, we get a reduced density matrix:
\begin{equation}
\rho = 
\frac{\text{Tr}_B \left( \ket{\mathcal{L}}\bra{\mathcal{L}} \right)}
{\bra{\mathcal{L}}\ket{\mathcal{L}}} \equiv \frac{\sigma}{\text{Tr}[\sigma]} ~,
\end{equation}
where, for computational ease, we have defined an unnormalized matrix $\sigma \equiv \text{Tr}_B \left( \ket{\mathcal{L}}\bra{\mathcal{L}} \right)$. All the bi-partite entanglement measures associated with the link state can be obtained using matrix $\rho$ or $\sigma$. For example, the Rényi entropy of index $m$ can be calculated as:
\begin{equation}
\mathcal{R}_m = \frac{\ln\left[\text{Tr}\,\rho^m\right]}{1-m} = \frac{1}{1-m}\, \ln \left[ \frac{\text{Tr}\, \sigma^m}{(\text{Tr} \, \sigma)^m} \right] ~.
\end{equation}
The entanglement entropy can be obtained by taking the $m\to 1$ limit of the Rényi entropy:
\begin{equation}
\text{EE} = \lim_{m \to 1} \mathcal{R}_m ~.
\end{equation}
\subsection{$T_{d,\,dn}$ torus links and associated quantum state}
A $T_{d,\,dn}$ torus link consists of $d$ copies of $T_{1,\,n}$ knots:
\begin{equation}
T_{d,\,dn} = \underbrace{T_{1,\,n}\, \sqcup \, T_{1,\,n} \, \sqcup \, \cdots \, \sqcup \, T_{1,\,n}}_{\text{$d$ copies}} ~,
\end{equation}
where each component $T_{1,n}$ is isotopic to the unknot and can be embedded on the surface of a torus so that it winds once around the meridian and $n$ times around the longitude. Moreover, every pair of distinct components has the same linking number $\text{lk}=n$. Thus, the $T_{d,\,dn}$ family consists of $d$ unknotted components with a uniform pairwise linking number $n$, making it a particularly simple and symmetric class of links.

For the $T_{d,\,dn}$ family, the Chern-Simons partition function on $S^3$ in the presence of a $T_{d,\,dn}$ link admits a remarkably simple closed-form expression in terms of the modular $\mathcal{S}$ and $\mathcal{T}$ matrices, which furnish a unitary representation of the modular group SL(2, $\mathbb{Z}$). Following Refs.~\cite{Stevan:2010jh,Brini:2011wi}, the partition function for the SU(2) group is: 
\begin{equation}
 Z(S^3;\,T_{d,\,dn}[a_1,a_2,\ldots,a_d]) = \sum_{b=0}^k \frac{(\mathcal{S} \mathcal{T}^n \mathcal{S})_{b0} }{(\mathcal{S}_{0b})^{d-1}} \left(\mathcal{S}_{a_1 b} \, \mathcal{S}_{a_2 b} \ldots \mathcal{S}_{a_d b} \right) ~,
 \label{TorusLinkPF}
\end{equation}
where the modular matrix elements for the SU(2) group are:
\begin{equation}
\mathcal S_{ab}
=
\sqrt{\frac{2}{k+2}}\,
\sin\left[
\frac{(a+1)(b+1)\pi}{k+2}
\right],
\qquad
\mathcal T_{ab}
=
\exp\left[
\frac{i\pi a(a+2)}{2(k+2)}
-\frac{i\pi k}{4(k+2)}
\right]\delta_{ab}.
\end{equation}
For convenience, we introduce $K\equiv k+2$. Since $K/k\to1$ in the large-$k$ limit, we will use \(K\) and \(k\) interchangeably in algebraic prefactors and power-law asymptotic estimates whenever their difference is subleading. However, this replacement must be handled with care in the oscillatory phases arising from the modular $\mathcal{T}$ matrix elements. In particular, the Rényi entropies and their large-level limits can depend on the level modulo $n$, making the distinction between $k$ modulo $n$ and $k+2$ modulo $n$ important. We therefore use $K$ as a bookkeeping device to preserve the exact phase information, keeping the distinction between $k$ and $K$ explicit in the phase factors whenever necessary, while suppressing it elsewhere when it does not affect the leading asymptotics.

Substituting the partition function \eqref{TorusLinkPF} into \eqref{LinkComplstate}, the corresponding link complement state becomes:
\begin{equation}
\ket{T_{d,\,dn}} = \sum_{a_1,\,\ldots,\,a_d=0}^k \,\sum_{b=0}^k \frac{(\mathcal{S} \mathcal{T}^n \mathcal{S})_{b0} }{(\mathcal{S}_{0b})^{d-1}} \left(\mathcal{S}_{a_1 b} \, \mathcal{S}_{a_2 b} \ldots \mathcal{S}_{a_d b} \right)\, \ket{e_{a_1},e_{a_2},\ldots,e_{a_d}} ~.
\label{}
\end{equation}
A more convenient representation of the state is obtained by performing the same local unitary transformation on each copy of the Hilbert space. Since local unitary transformations preserve all entanglement measures, they do not alter the entanglement properties of the state. Using the unitarity of the modular $\mathcal{S}$ matrix, we define a new basis:
\begin{equation}
\ket{e_{a_i}} = \sum_{c_i=0}^k (\mathcal{S}_{a_i \, c_i})^* \ket{f_{c_i}}
\end{equation}
for the $i^{\mathrm{th}}$ Hilbert space. Using the symmetric property of $\mathcal{S}$ matrix and employing the unitarity relation
\begin{equation}
\sum_x
\mathcal S_{xy}\, \mathcal S_{xz}^{*} = \delta_{yz} ~,
\end{equation}
the sum over the representation labels collapses immediately, and the state assumes the GHZ-like form:
\begin{equation}
\ket{T_{d,\,dn}} = \sum_{R=0}^k  \frac{(\mathcal{S} \mathcal{T}^{n} \mathcal{S})_{R0}}{(\mathcal{S}_{0R})^{d-1}} \, \ket{f_{R},\,f_{R},\,\ldots,\,f_{R}} ~.
\label{LinkstateCompact}
\end{equation}
The above GHZ-like structure of the state tells us that the total density matrix is a diagonal matrix of order $(k+1)^d$, which has only $(k+1)$ non-zero diagonal elements. This diagonal structure ensures that the reduced density matrix $\rho$ obtained by any bi-partitioning of the total Hilbert space remains diagonal. Since the diagonal entries are also the eigenvalues of $\rho$, and the zero eigenvalues do not contribute to the entanglement measures, the entanglement structure is insensitive to how many parties are traced out. So without loss of generality, we can assume the bipartition $(A|B) = (1|d-1)$. Tracing out the Bob system will give the reduced density matrix $\rho$ of order $(k+1)$, where the eigenvalues are labeled by the integrable representations, given as follows:
\begin{equation}
\lambda_R = \frac{\abs{F_R}^2}{\abs{F_0}^2 + \abs{F_1}^2 + \ldots + \abs{F_k}^2} \quad;\quad F_{R} =
\frac{\left(\mathcal{S}\mathcal{T}^{n}\mathcal{S}\right)_{R 0}}{(\mathcal{S}_{0 R})^{d-1}} ~.
\label{FuncFR}
\end{equation}
For our calculation, we define the unnormalized matrix $\sigma$ such that $\rho = \sigma/\text{Tr}[\sigma]$. The unnormalized eigenvalues $\Lambda_R$ of matrix $\sigma$ are given as:
\begin{equation}
\Lambda_R = \abs{F_R}^2 ~.
\end{equation}
\section{Large-$k$ asymptotics of the density matrix moments} \label{sec3}
In this section, we investigate the large-\(k\) asymptotics of the density-matrix moments for the \(T_{d,dn}\) torus-link states introduced in Section~2.3. Our analysis focuses on the SU(2)$_k$ Chern--Simons theory and exploits the modular-matrix representation of the corresponding link-state coefficients. We organize the asymptotic analysis into contributions from the left edge, bulk, and right edge of the integrable-representation space, and identify the leading contribution to the moments in the large-\(k\) limit.

As discussed earlier, the Rényi entropies can be expressed in terms of the moments of the matrix \(\sigma\), which is the unnormalized version of the reduced density matrix. For the present class of states, these moments take the form
\begin{equation}
\text{Tr}[\sigma^m] = \sum_{R = 0}^{k} (\Lambda_{R})^m = \sum_{R = 0}^{k} |F_{R}|^{2m} ~.
\end{equation}
To determine the large-\(k\) behavior of the entanglement measures, we therefore need to understand the asymptotic behavior of these moments. To begin with, consider the function \(F_R\) defined in Eq.~\eqref{FuncFR}. Using the fact that the modular \(\mathcal T\)-matrix is diagonal, it can be written as
\begin{equation}
F_R
=
\frac{1}{(\mathcal S_{0R})^{d-1}}
\sum_{a=0}^{k} \mathcal S_{Ra}\,
\mathcal T_{aa}^{\,n}\,
\mathcal S_{a0} ~.
\label{FuncFRSum}
\end{equation}
There is an intermediate sum over the variable $a$ in the function $F_R$ written in this form. Let us first analyze the large-\(k\) or large-$K$ behavior of this sum. In this limit, the summand takes the leading-order form:
\begin{equation}
\mathcal S_{Ra}\,
\mathcal T_{aa}^{\,n}\,
\mathcal S_{a0} \, \longrightarrow \, G_K(a/K,\, R) = \frac{\sqrt{2}\, \sin[\pi  (R+1) \frac{a}{K} ]}{\sqrt{K}} \, e^{in \pi \left(\frac{K}{2}\frac{a^2}{K^2} + \frac{a}{K}-\frac{1}{4}\right)} \, \frac{\sqrt{2}\, \sin[\pi \frac{a}{K} ]}{\sqrt{K}} ~,
\label{GKFunction}
\end{equation}
where we have explicitly written the function $G$ in the variable $a/K$ to enable us to apply the Poisson summation formula, which we review below.
\begin{reviewbox}{Review: Poisson Summation Formula}
For a sufficiently regular function \(f\) and its Fourier transform \(\widehat f\),
\[
\sum_{m\in\mathbb Z}f(m)
=
\sum_{s\in\mathbb Z}\widehat f(s)
=
\sum_{s\in\mathbb Z}
\int_{-\infty}^{\infty}
f(y)e^{-2\pi i s y}\,dy.
\]
\end{reviewbox}
Applying the Poisson summation formula for the function $G$, we will have:
\begin{equation}
\sum_{a\,\in\,\mathbb Z} G_K(a/K,\,R) = \sum_{s\,\in\,\mathbb Z}\, \int_{-\infty}^\infty G_K(y/K,\,R)\, e^{-2\pi i s y}\, dy  ~.
\end{equation}
However, in \eqref{FuncFRSum}, the sum over $a$ is restricted to integers from $0$ to $k$. To apply the Poisson summation formula to the finite sum, we extend $G$ from integer values of \(a\) to a sufficiently smooth function of a continuous variable \(y\), and introduce a smooth cutoff function \(\chi(y/K)\) that equals unity on the interval \(0\leq y/K\leq1\) and has compact support in a slightly larger interval. The finite sum can then be represented by the Poisson-resummed expression
\begin{equation}
\sum_{a=0}^{k} G_K(a/K,\,R)
=
\sum_{s\in\mathbb Z}
\int_{-\infty}^{\infty}
G_K(y/K,\,R)\,
\chi(y/K)\,
e^{-2\pi i s y}\,dy ~,
\end{equation}
where the cutoff ensures that only the desired range of the summation contributes. Next, we invoke the variable change $x=y/K$ in the integral on the right-hand side. Now the cut-off function $\chi(x)$ equals unity on the interval $0 \leq x \leq 1$ and smoothly decays to zero outside this interval. Since the cutoff is unity throughout the original integration domain and suppresses contributions away from it, we may effectively restrict the resulting integral to $[0,1]$, giving the final resummed expression:
\begin{equation}
\sum_{a=0}^k G_K(a/K,\,R) = K\,\sum_{s\,\in\,\mathbb Z}\, \int_{0}^1 G_K(x,\,R)\, e^{-2\pi i K s x}\, dx  ~.
\label{GDiscreteToCont}
\end{equation}
Thus, we arrive at the expression: 
\begin{equation}
F_{R}
\,\, \sim \,\, \frac{K}{\left(\mathcal{S}_{0R}\right)^{d-1}}  \,\sum_{s\,\in\,\mathbb Z}\, \int_{0}^1 G_K(x,\,R)\, e^{-2\pi i K s x}\, dx ~.
\label{FRFunctionAfteraSum}
\end{equation}
Since the moments $\text{Tr}[\sigma^m]$ require summing $|F_R|^{2m}$ over various values of $R$, we need to understand the asymptotic behavior of $F_R$ for various values of $R$. 

\subsection{Asymptotic framework and decomposition of the representation space}
A key feature of the \(T_{d,dn}\) family is that the asymptotic contributions to \(\operatorname{Tr}[\sigma^m]\) can be organized into three distinct regions of the representation space. To characterize the different asymptotic regimes, we introduce the rescaled variable
\[
r\equiv\frac{R}{K} ~,
\]
with $K=k+2$. Since \(R=0,1,\ldots,k\), the variable \(r\) can be treated as a continuous variable in the large-\(k\) limit, with \(r\in[0,1]\). We distinguish three regimes according to how \(R\) scales with \(k\). The `left-edge region' corresponds to \(R=O(1)\), for which \(r\to 0\) as \(k\to\infty\). Similarly, the `right-edge region' corresponds to \(k-R=O(1)\), for which \(r\to1\). The remaining representations, for which \(R=O(k)\) and \(r\) remains finite in the open interval \(0<r<1\), constitute the `bulk region'.

A central observation of this work is that the eigenvalues associated with the two edge regions provide the leading contribution to \(\operatorname{Tr}[\sigma^m]\), whereas the contribution from the bulk region is asymptotically subleading. Accordingly, we decompose the moments as
\begin{equation}
\text{Tr}[\sigma^m] = \text{Tr}[\sigma^m]_{\text{left}} + \text{Tr}[\sigma^m]_{\text{bulk}} + \text{Tr}[\sigma^m]_{\text{right}}   ~,
\end{equation}
where the contribution $\text{Tr}[\sigma^m]_{\text{bulk}}$ is subleading compared to $\text{Tr}[\sigma^m]_{\text{left}}$ and $\text{Tr}[\sigma^m]_{\text{right}}$.
In the following, we use the method of stationary phase to evaluate the contributions from these three regions and thereby determine the large-\(k\) asymptotics of \(\operatorname{Tr}[\sigma^m]\).

\subsection{Left-edge contribution}
The left-edge region corresponds to $R=O(1)$ representations such that $r=R/K \to 0$. So we set $R = \alpha$, where $\alpha$ is held fixed as $k \to \infty$. In the large-$k$ limit, up to leading order, we have:   
\begin{equation}
\mathcal{S}_{0R} = \mathcal{S}_{0 \alpha} = \sqrt{\frac{2}{k+2}}\,
\sin\left[
\frac{(\alpha+1)\pi}{k+2}
\right] \, \sim \, 
\frac{\sqrt{2}\, \pi (\alpha+1)}{K^{3/2}}  ~.
\end{equation}
Furthermore, using \eqref{GKFunction}, the leading form of $G_K(x,R)$ in the left-edge region is
\begin{equation}
G_K(x,\alpha) \, \sim \, \frac{2}{K} \,\sin [\pi  x]\, \sin[\pi  (\alpha +1) x]\, \exp[i K \left( \frac{n \pi x^2}{2} + \frac{n \pi x}{K} - \frac{n \pi}{4K} \right) ] ~.
\label{}
\end{equation}
Note that we have written the phase in the form $e^{iK(\cdots)}$ to make explicit the large parameter ($K$) governing the oscillatory integral. This form is convenient for applying the method of the stationary phase in the subsequent analysis. Substituting the leading forms of $\mathcal{S}_{0R}$ and $G_K(x,R)$ into \eqref{FRFunctionAfteraSum}, we get $F_R$ in the left-edge region as:
\begin{equation}
F_\alpha = \frac{K^{(3d-3)/2}}{ 2^{\frac{d-3}{2}}\,  \pi^{d-1}\, (\alpha +1)^{d-1} } \, \sum_{s\,\in\,\mathbb Z}\, \int_{0}^1 \sin [\pi  x]\, \sin[\pi  (\alpha +1) x]\, e^{i K\, \Phi_s(x)} \, dx ~.
\label{}
\end{equation}
where the phase function is 
\begin{equation}
\Phi_s(x) = \frac{n \pi x^2}{2} + \frac{n \pi x}{K} - \frac{n \pi}{4K} -2\pi s x ~.
\end{equation}
For determining the leading-order stationary points in the large-$k$ limit, the terms suppressed by $1/K$ in $\Phi_s(x)$ can be neglected. The resulting leading-order phase is therefore
\begin{equation}
\Phi_s(x) = \frac{n \pi x^2}{2} - 2\pi s x .
\end{equation}
Thus,
\begin{equation}
F_\alpha = \frac{K^{(3d-3)/2}}{ 2^{\frac{d-3}{2}}\,  \pi^{d-1}\, (\alpha +1)^{d-1} } \, \sum_{s\,\in\,\mathbb Z}\, I_s ~,
\label{Funalpha}
\end{equation}
where the integral is given as:
\begin{equation}
I_s = \int_{0}^1 \sin [\pi  x]\, \sin[\pi  (\alpha +1) x]\, e^{i K\, \Phi_s(x)}\, dx ~.
\label{IntegralIs}
\end{equation}
The large-$k$ asymptotics of these integrals can then be obtained using the method of stationary phase, which we briefly review below.
\begin{reviewbox}{Review: Stationary phase approximation}
The stationary phase approximation \cite{Copson_1965} is used to perform asymptotic analysis of functions obtained by integrating against a rapidly varying complex exponential.
Consider the integral:
\begin{equation}
I = \int_{a}^b f(x)\, e^{iK\, \phi(x)}\, dx ~,
\end{equation}
where $K$ is a large parameter. The functions $f(x)$ and $\phi(x)$ are smooth and $\phi'(x)$ has only a finite number of zeros in the interval $[a,b]$. In other words, the function $\phi(x)$ has finitely many stationary points in $[a,b]$. The method of the stationary phase tells us that the leading order contribution in the large-$K$ expansion of $I$ will come from the immediate neighborhoods of stationary points in $[a,b]$ of the phase function $\phi(x)$. Let $x_0$ be a stationary point in the interval $[a,b]$. Let us assume that $\phi''(x_0)>0$, which is applicable for our calculation. In the neighborhood of $x_0$, we have Taylor expansions:
\begin{equation}
f(x) = \sum_{j=0}^{\infty}\frac{f^{(j)}(x_0)}{j!}(x-x_0)^j \quad;\quad \phi(x)=\phi(x_0)+ \frac{\phi''(x_0)}{2!}(x-x_0)^2 + \ldots ~,
\end{equation}
where we ignore the higher-order derivative terms in the expansion of $\phi(x)$. Thus, we can approximate the integral about a small neighborhood around $x_0$ as:
\begin{equation}
I \, \sim \, \int_{\text{nbh}(x_0)}\, \sum_{j=0}^{\infty}\frac{f^{(j)}(x_0)}{j!}(x-x_0)^j\, e^{iK \phi(x_0)} \,e^{iK \frac{\phi''(x_0)}{2}(x-x_0)^2} \, dx ~.
\end{equation}
Here, the neighborhood $\text{nbh}(x_0)$ will depend on the location of the stationary point. If $x_0 = a$, then $\text{nbh}(a)=[a,\,a+\epsilon]$ for some small positive number $\epsilon$. If $x_0 = b$, then $\text{nbh}(b)=[b-\epsilon,\,b]$. For $a < x_0 < b$, we have $\text{nbh}(x_0)=[x_0-\epsilon,\,x_0+\epsilon]$. Next, introduce a change of variable $u=(x-x_0)\sqrt{K\, \phi''(x_0)/2}$.
\begin{equation}
I \, \sim \, \frac{\sqrt{2}\,e^{iK \phi(x_0)}}{\sqrt{K\,\phi''(x_0)}}  \sum_{j=0}^{\infty} \left[ A(j)\,  \frac{2^{j/2}\,f^{(j)}(x_0)}{j!\,(K\, \phi''(x_0))^{j/2}} \right]  ~.
\label{Istationaryphase}
\end{equation}
where the function $A(j)$ is generalization of the standard Fresnel integrals:
\begin{equation}
A(j) = \int_{\text{range}(u)}\, u^j \,e^{iu^2} \, du     
\end{equation}
Here $\text{range}(u)$ is the range of the integration variable $u$ and will again depend on the location of the stationary point. If $x_0 = a$, then $\text{range}(u)=[0,\infty]$. If $x_0 = b$, then $\text{range}(u)=[-\infty,0]$. For $a < x_0 < b$, we have $\text{range}(u)=[-\infty,\infty]$. Since $A(j)$ is independent of $K$, the \eqref{Istationaryphase} provides the large-$K$ expansion of $I$ to all orders in $1/K$ in the neighborhood of the stationary point $x_0$. If $\phi(x)$ has multiple isolated stationary points in the interval $[a,b]$, the contributions from all such stationary points must be summed to obtain the complete large-$K$ asymptotic expansion of $I$. 
\end{reviewbox}
Let us now apply the stationary phase analysis to the integral $I_s$ in \eqref{IntegralIs}. We identify
\begin{equation}
f(x) = \sin [\pi  x]\, \sin[\pi  (\alpha +1) x] \quad;\quad \phi(x) = \frac{n \pi \, x^2}{2} - 2\pi s x ~.
\end{equation}
The stationary points $x_0 \in [0,1]$ are determined by $\phi'(x_0) = 0$ giving
\begin{equation}
x_0=\frac{2s}{n} ~.
\end{equation}
Since $s \in \mathbb{Z}$, the condition $0 \leq x_0 \leq 1$ restricts the the allowed values of $s$ to 
\begin{equation}
s= \begin{cases}
    0,1,2,\ldots,\dfrac{n}{2} & \text{,\,\, for $n=$ even} \\[0.5cm] 0,1,2,\ldots,\dfrac{n-1}{2} & \text{,\,\, for $n=$ odd}
\end{cases}  
\end{equation}
The corresponding stationary points are
\begin{equation}
x_0 = \begin{cases}
    0,\dfrac{2}{n},\dfrac{4}{n},\ldots,\dfrac{n-2}{n},1 & \text{,\,\, for $n=$ even} \\[0.5cm] 0,\dfrac{2}{n},\dfrac{4}{n},\ldots,\dfrac{n-3}{n},\dfrac{n-1}{n} & \text{,\,\, for $n=$ odd}
\end{cases} 
\end{equation}
At the two endpoints $x_0 = 0$ and $x_0 = 1$, both $f(x_0)$ and $f'(x_0)$ vanish. Consequently, the corresponding endpoint stationary-point contributions begin at order \(K^{-3/2}\), whereas the interior stationary points contribute at order \(K^{-1/2}\). The leading contribution, therefore, comes only from the stationary points satisfying \(0<x_0<1\). These stationary points are given by $x_0 = 2s/n$ where $s$ takes the following values:
\begin{equation}
s= 1,2,3,\ldots, \left\lfloor \frac{n-1}{2}\right\rfloor ~.
\end{equation}
For these stationary points, we have:
\begin{equation}
f(x_0) =  \sin \left[\frac{2 \pi  s}{n}\right] \sin\left[\frac{2 \pi  (\alpha +1) s}{n}\right] \quad;\quad \phi(x_0) = -\frac{2 \pi  s^2}{n} \quad;\quad \phi''(x_0) = n \pi ~.
\end{equation}
Applying the stationary phase formula \eqref{Istationaryphase} to the integral $I_s$ in \eqref{IntegralIs}, we get, at leading order:
\begin{equation}
I_s  \, \sim \,   
\frac{\sqrt{2}\, e^{-\frac{2 i \pi K s^2}{n}}}{\sqrt{K\,n \pi }} \, \sin\left[\frac{2 \pi  s}{n}\right] \sin\left[\frac{2 \pi  (\alpha +1) s}{n}\right] A(0) ~,
\end{equation}
where the integral $A(0)$ is given as:
\begin{equation}
A(0) =  \int_{-\infty}^{\infty} \,e^{iu^2} \, du  =  e^{\frac{i \pi }{4}} \sqrt{\pi } ~.
\end{equation}
The large-$k$ asymptotic of $F_{\alpha}$ in \eqref{Funalpha} will be:
\begin{equation}
F_\alpha \, \sim \,   \frac{e^{\frac{i \pi }{4}}\, K^{(3d-4)/2}}{2^{\frac{d-4}{2}}\,n^{1/2}\,  \pi^{d-1}\, (\alpha +1)^{d-1} } \, \sum_{s=1}^{\left\lfloor \frac{n-1}{2}\right\rfloor} \, e^{-\frac{2 i \pi K s^2}{n}} \, \sin\left[\frac{2 \pi  s}{n}\right] \sin\left[\frac{2 \pi  (\alpha +1) s}{n}\right]  ~.
\label{}
\end{equation}
Finally, we arrive at the leading order large-$k$ asymptotic of the corresponding eigenvalues $\Lambda_R = |F_R|^2$ in the left-edge region:
\begin{equation}
\Lambda_\alpha \, \sim \, \frac{K^{3d-4}}{2^{d-4}\,n\,  \pi^{2d-2}\, (\alpha +1)^{2d-2} }  \,\left| \sum_{s=1}^{\left\lfloor \frac{n-1}{2}\right\rfloor}\, e^{-\frac{2 i \pi  K s^2}{n}} \sin\left[\frac{2 \pi  s}{n}\right] \sin\left[\frac{2 \pi  (\alpha +1) s}{n}\right]  \right|^2 ~.
\label{LeftEigenAsymp}
\end{equation}
We can now write the leading contribution from the left-edge region to $\text{Tr}[\sigma^m]$ as
\begin{equation}
\text{Tr}[\sigma^m]_{\text{left}} = \sum_{\alpha=0}^{\infty} (\Lambda_\alpha)^m ~,
\end{equation}
where the large-$K$ asymptotic form of $\Lambda_\alpha$ is given in \eqref{LeftEigenAsymp}. At finite \(K\), the sum over \(\alpha\) terminates at \(k=K-2\). In the left-edge scaling limit, \(\alpha\) is held fixed as \(K\to\infty\), and the upper limit consequently tends to infinity. Thus, provided the resulting large-\(\alpha\) series is convergent, the limiting left-edge contribution is represented by the infinite sum above.

To carry out the sum over $\alpha$, we first note from \eqref{LeftEigenAsymp} that the \(\alpha\)-dependent oscillatory factor contains  
\[
\sin\left[\frac{2\pi(\alpha+1)s}{n}\right] ~.
\]
Since $s$ is an integer, this factor is periodic in $\alpha$ with period $n$. We can therefore decompose the non-negative integer \(\alpha\) as 
\[
\alpha=np+q,\qquad p\in\mathbb Z_{\geq0},\quad q=0,1,\ldots,n-1.
\]
This parametrization covers all non-negative integers \(\alpha\) exactly once. Hence, we can rewrite the left-edge contribution as:
\begin{equation}
\text{Tr}[\sigma^m]_{\text{left}} = \sum_{p=0}^{\infty}\, \sum_{q=0}^{n-1} \,(\Lambda_{np+q})^m  ~.
\label{Trsigmaleftsum}
\end{equation}
The advantage of this decomposition is that the oscillatory factor in \eqref{LeftEigenAsymp} becomes independent of \(p\):
\[
\sin\left[\frac{2 \pi  (\alpha +1) s}{n}\right] = \sin\left[\frac{2 \pi  (np+q +1) s}{n}\right] = \sin\left[\frac{2 \pi  (q+1) s}{n}\right] ~.
\]
The remaining \(p\)-dependence can therefore be summed independently. In particular,
\begin{equation}
\sum_{p=0}^{\infty}\, \frac{1}{(np+q+1)^{2dm-2m}}  = n^{2m-2 dm}\, \zeta\left(2 dm-2m,\,\dfrac{q+1}{n}\right) ~,
\end{equation}
where $\zeta\left(s,\,a\right)$ denotes the Hurwitz zeta function. Since our set-up has $d\geq 2$ and $m\geq 1$, the series converges because $2dm-2m > 1$. Performing the sum over $p$, we obtain
\begin{equation}
\text{Tr}[\sigma^m]_{\text{left}} = \frac{K^{3dm-4m}}{2^{dm-4m}\,\pi^{2dm-2m}\,n^{2dm-m}}  \,\sum_{q=0}^{n-1} \,X(q)^m\, \zeta\left(2dm-2m,\,\frac{q+1}{n}\right)  ~,
\label{TrSigmaLeftFinal}
\end{equation}
where we have defined:
\begin{equation}
X(q) = \left| \sum_{s=1}^{\left\lfloor \frac{n-1}{2}\right\rfloor}\, e^{-\frac{2 i \pi  K s^2}{n}} \sin\left[\frac{2 \pi  s}{n}\right] \sin\left[\frac{2 \pi  (q +1) s}{n}\right]  \right|^2  ~.
\label{FuncXq}
\end{equation}
Thus, the left-edge contribution scales as 
\begin{equation}
\operatorname{Tr}[\sigma^m]_{\text{left}}
\, \sim \, k^{3dm-4m} ~,
\end{equation}
with the leading coefficient given by the expression in \eqref{TrSigmaLeftFinal}.
We have thus obtained the leading contribution from the left-edge region, \(R=O(1)\). We next analyze the opposite boundary of the representation space, corresponding to \(k-R=O(1)\), and determine the associated right-edge contribution to \(\operatorname{Tr}[\sigma^m]\).
\subsection{Right-edge contribution}
The right-edge region consists of representations $R$ such that $k-R=O(1)$, corresponding to $r=R/K \to 1$. We therefore set $R = k- \beta$, where $\beta$ is held fixed as $k \to \infty$. It should be noted that the modular $\mathcal{S}$ matrix has symmetry that $\mathcal{S}_{0,\,{k-\beta}} = \mathcal{S}_{0 \beta}$. Therefore, in the large-$k$ limit, up to leading order, we have:   
\begin{equation}
\mathcal{S}_{0R} = \mathcal{S}_{0,\,{k-\beta}} = \mathcal{S}_{0 \beta}  \, \sim \, 
\frac{\sqrt{2}\, \pi (\beta+1)}{K^{3/2}}  ~.
\end{equation}
Further, note the following important property in the right-edge region:
\begin{equation}
\mathcal{S}_{Ra} = \mathcal{S}_{k-\beta,\,a} = (-1)^a\, \mathcal{S}_{\beta a} ~.
\end{equation}
Therefore, for the right-edge region  
\begin{equation}
\mathcal S_{Ra}\,
\mathcal T_{aa}^{\,n}\,
\mathcal S_{a0}  = (-1)^a\, \mathcal{S}_{\beta a} \,
\mathcal T_{aa}^{\,n}\,
\mathcal S_{a0} \, \sim \, (-1)^a \,\frac{\sqrt{2}\, \sin[\pi  (\beta+1) \frac{a}{K} ]}{\sqrt{K}} \, e^{in \pi \left(\frac{K}{2}\frac{a^2}{K^2} + \frac{a}{K}-\frac{1}{4}\right)} \, \frac{\sqrt{2}\, \sin[\pi \frac{a}{K} ]}{\sqrt{K}} ~.
\label{}
\end{equation}
So, we extract the leading form of $G_K(x,R)$ in \eqref{GKFunction} with $x=a/K$, for the right-edge region as:
\begin{equation}
G_K(x,k-\beta) \, \sim \, \frac{2}{K} \,\sin [\pi  x]\, \sin[\pi  (\beta +1) x]\, \exp[i K \left( \frac{n \pi x^2}{2} + \frac{n \pi x}{K} - \frac{n \pi}{4K} + \pi x \right) ] ~.
\label{}
\end{equation}
The factor \((-1)^a=e^{i\pi a}=e^{iK\pi x}\) therefore introduces an additional phase term \(e^{iK\pi x}\) in the above expression. Substituting these leading order forms into \eqref{FRFunctionAfteraSum}, we get $F_R$ in the right-edge region as:
\begin{equation}
F_{k-\beta} = \frac{K^{(3d-3)/2}}{ 2^{\frac{d-3}{2}}\,  \pi^{d-1}\, (\beta +1)^{d-1} } \, \sum_{t\,\in\,\mathbb Z}\, \int_{0}^1 \sin [\pi  x]\, \sin[\pi  (\beta +1) x]\, e^{i K\, \Phi_t(x)} \, dx ~.
\label{}
\end{equation}
where the phase function, after neglecting the terms suppressed by $1/K$, which do not contribute to determining the leading-order stationary points, is 
\begin{equation}
\Phi_t(x) = \frac{n \pi x^2}{2} + \pi x - 2\pi t x  ~.
\end{equation}
Thus,
\begin{equation}
F_{k-\beta} = \frac{K^{(3d-3)/2}}{ 2^{\frac{d-3}{2}}\,  \pi^{d-1}\, (\beta +1)^{d-1} } \, \sum_{t\,\in\,\mathbb Z}\, I_t ~,
\label{Funbeta}
\end{equation}
where the integral is given as:
\begin{equation}
I_t = \int_{0}^1 \sin [\pi  x]\, \sin[\pi  (\beta +1) x]\, e^{i K\, \Phi_t(x)}\, dx ~.
\label{IntegralIst}
\end{equation}
We now apply the stationary phase analysis to $I_t$, with
\begin{equation}
f(x) = \sin [\pi  x]\, \sin[\pi  (\beta +1) x] \quad;\quad \phi(x) = \frac{n \pi \, x^2}{2} + \pi x - 2\pi t x ~.
\end{equation}
The stationary points are determined by $\phi'(x_0) = 0$, giving
\begin{equation}
x_0=\frac{2t-1}{n} ~.
\end{equation}
As in the left-edge analysis, an interior stationary point gives a contribution of order \(K^{-1/2}\), whereas an endpoint stationary point for which both \(f(x_0)\) and \(f'(x_0)\) vanish contributes only at order \(K^{-3/2}\). We therefore retain only the stationary points satisfying \(0<x_0<1\). This gives
\begin{equation}
t= 1,2,3,\ldots, \left\lfloor \frac{n}{2}\right\rfloor ~.
\end{equation}
The corresponding stationary points will be $x_0 = (2t-1)/n$, for which we will have:
\begin{equation}
f(x_0) =  \sin \left[\frac{\pi(2t-1)}{n}\right] \sin\left[\frac{ \pi  (\beta +1) (2t-1)}{n}\right] \quad;\quad \phi(x_0) = -\frac{ \pi(2t-1)^2}{2n} \quad;\quad \phi''(x_0) = n \pi ~.
\end{equation}
Therefore, using \eqref{Istationaryphase}, the leading large-\(K\) asymptotics of \(I_t\) is
\begin{equation}
I_t  \, \sim \,   
\frac{\sqrt{2}\, e^{-\frac{i \pi  K (2t-1)^2}{2n}}}{\sqrt{K\,n \pi }} \, e^{\frac{i \pi }{4}} \sqrt{\pi } \, \sin \left[\frac{\pi(2t-1)}{n}\right] \sin\left[\frac{ \pi  (\beta +1) (2t-1)}{n}\right] ~.
\end{equation}
Finally, using \eqref{Funbeta}, the leading large-$k$ asymptotic of the eigenvalues $\Lambda_{k-\beta} = |F_{k-\beta}|^2$ in the right-edge region is
\begin{equation}
\Lambda_{k-\beta} \, \sim \, \frac{K^{3d-4}}{2^{d-4}\,n\,  \pi^{2d-2}\, (\beta +1)^{2d-2} }  \,\left| \sum_{t=1}^{\left\lfloor \frac{n}{2}\right\rfloor}\, e^{-\frac{i \pi  K (2t-1)^2}{2n}} \sin\left[\frac{\pi(2t-1)}{n}\right] \sin\left[\frac{\pi  (\beta +1) (2t-1)}{n}\right]  \right|^2 ~.
\label{RightEigenAsymp}
\end{equation}
The leading contribution from the right-edge region to $\text{Tr}[\sigma^m]$ can therefore be written as
\begin{equation}
\text{Tr}[\sigma^m]_{\text{right}} = \sum_{\beta=0}^{\infty} (\Lambda_{k-\beta})^m ~,
\end{equation}
where, as in the left-edge analysis, the infinite sum represents the \(K\to\infty\) limit of the finite sum over \(0\leq\beta\leq k\), with \(\beta\) held fixed in the right-edge scaling regime. We now use the same decomposition as in the left-edge analysis, 
\[
\beta=np+q,\qquad p\in\mathbb Z_{\geq0},\quad q=0,1,\ldots,n-1 ~.
\]
This parametrization covers all non-negative integers \(\beta\) exactly once. Moreover, 
\[
\sin\left[\frac{\pi(\beta +1)(2t-1)}{n}\right] = \sin\left[\frac{\pi(np+q +1)(2t-1)}{n}\right] = (-1)^p\,\sin\left[\frac{\pi(q +1)(2t-1)}{n}\right] ~.
\]
Since the factor \((-1)^p\) is independent of \(t\), it factors out of the sum and drops out upon taking the modulus squared in \eqref{RightEigenAsymp}. The remaining \(p\)-dependence is therefore entirely contained in the algebraic factor $(np+q+1)^{-2d+2}$ , and its sum gives the same Hurwitz zeta function encountered in the left-edge analysis. We consequently obtain
\begin{equation}
\text{Tr}[\sigma^m]_{\text{right}} = \frac{K^{3dm-4m}}{2^{dm-4m}\,\pi^{2dm-2m}\,n^{2dm-m}}  \,
\sum_{q=0}^{n-1} \,Y(q)^m\, \zeta\left(2dm-2m,\,\frac{q+1}{n}\right) ~,
\label{TrSigmaRightFinal}
\end{equation}
where
\begin{equation}
Y(q) = \left| \sum_{t=1}^{\left\lfloor \frac{n}{2}\right\rfloor}\, e^{-\frac{i \pi K (2t-1)^2}{2n}} \sin\left[\frac{\pi(2t-1)}{n}\right] \sin\left[\frac{\pi  (q +1) (2t-1)}{n}\right]  \right|^2  ~.
\label{FuncYq}
\end{equation}
Thus, the right-edge contribution scales as 
\begin{equation}
\operatorname{Tr}[\sigma^m]_{\text{right}}
\, \sim \, K^{3dm-4m} ~,
\end{equation}
with the leading coefficient given by \eqref{TrSigmaRightFinal}.

An interesting feature of the result is that, for odd \(n\), the left-edge and right-edge contributions to \(\operatorname{Tr}[\sigma^m]\) are identical. This follows from the symmetry $\Lambda_R = \Lambda_{k-R}$ for odd $n$, which exchanges the left and right edges of the representation space and explains the equality of their contributions for odd \(n\). This translates into an identical leading order asymptotic contribution in left-edge and right-edge contributions. We provide a proof for this below.
\begin{reviewbox}{Proof that $\text{Tr}[\sigma^m]_{\text{left}}=\text{Tr}[\sigma^m]_{\text{right}}$ for odd values of $n$}
The values of $\text{Tr}[\sigma^m]_{\text{left}}$ and $\text{Tr}[\sigma^m]_{\text{right}}$ are given in \eqref{TrSigmaLeftFinal} and \eqref{TrSigmaRightFinal} respectively. All terms are the same except the functions $X(q)$ and $Y(q)$ in the two cases, respectively. These functions are given in \eqref{FuncXq} and \eqref{FuncYq} respectively, and we need to prove that $X(q)=Y(q)$ for odd values of $n$. Take $n=2w+1$ with $w\in\mathbb{Z}$. So the upper limit of the sum in both functions becomes $w$, and both summations contain the exact same number of terms. In the function $Y$, we change the summation variable $t\rightarrow s$, where $s=w+1-t$. Since $t$ runs from 1 to $w$, the new variable $s$ takes values from $w$ to 1. The two sine terms in the function $Y$ will be modified in terms of the new summation variable as follows:
\begin{equation}
\sin\left[\frac{\pi(2t - 1)}{n}\right] \rightarrow \sin\left[\frac{2\pi s}{n}\right] \quad;\quad \sin\left[\frac{\pi(q + 1)(2t - 1)}{n}\right] \rightarrow (-1)^q\,\sin\left[\frac{2\pi(q + 1)s}{n}\right] ~.
\end{equation}
These are precisely the sine terms in the function $X(q)$ in \eqref{FuncXq} with additional phase $(-1)^q$. Since we need to take the modulus square after the summation, this overall phase will not matter, and we see that the trigonometric factors of $Y(q)$ map perfectly to those of $X(q)$. Let us now check the phase factors. The phase factor in $Y$ under the change of variable $t\rightarrow s$ where $s=w+1-t$ will change as follows:
\begin{equation}
\exp[-\frac{i \pi  k (2t-1)^2}{2n}] \rightarrow e^{-\frac{i \pi  k n}{2}} e^{2 i \pi  k s}\, \exp[-\frac{2 i \pi  k s^2}{n}] = e^{-\frac{i \pi  k n}{2}}\,\exp[-\frac{2 i \pi  k s^2}{n}] ~.
\end{equation}
This is again the same phase factor in $X(q)$ up to an overall phase of $e^{-\frac{i \pi  k n}{2}}$, which will not matter once we take the modulus square of the sum. Hence, the phase factors of $Y(q)$ also map perfectly to those of $X(q)$. Thus, $X(q)=Y(q)$ for odd values of $n$. Hence the proof.
\end{reviewbox}
We have thus determined the leading contributions to \(\operatorname{Tr}[\sigma^m]\) arising from the two edges of the representation space, corresponding to \(R=O(1)\) and \(k-R=O(1)\), respectively. It remains to examine the contribution from the bulk region, where \(R=O(k)\) and \(r=R/K\) remains finite in the interval \(0<r<1\). In the following subsection, we analyze the large-\(K\) behavior of the eigenvalues in this region and determine the corresponding contribution to \(\operatorname{Tr}[\sigma^m]\).
\subsection{Bulk contribution and its asymptotic suppression}
For the bulk region, $R=O(k)$ such that $r=R/K$ is finite in the interval $0<r<1$. We fix $r\in(0,1)$ and set
\[ R = \lfloor K r \rfloor = Kr - \{Kr\} ~, \]
where $\{Kr\}$ is the fractional part of $Kr$, satisfying $0 \leq \{Kr\} < 1$. In the large-$k$ limit with $r=R/K$ and $x=a/K$, the leading forms of the modular matrix elements are:
\begin{equation}
\mathcal{S}_{0R} \sim  \frac{\sqrt{2}\, \sin[\pi r]}{\sqrt{K}} \quad,\quad \mathcal{S}_{a0} \sim
\frac{\sqrt{2}\, \sin[\pi x]}{\sqrt{K}} \quad,\quad \mathcal{T}_{aa} \sim e^{i K \left(\frac{\pi  x}{K}-\frac{\pi }{4 K}+\frac{\pi  x^2}{2}\right)} ~.
\label{ModMatBulkAsymp}
\end{equation}
For the remaining modular matrix element, we have
\begin{equation}
\mathcal{S}_{R a} = \frac{\sqrt{2}\, \sin[\frac{\pi (a+1)(R+1)}{K}]}{\sqrt{K}} = \frac{\sqrt{2}\, \sin[\frac{\pi (Kx+1)(Kr - \{Kr\}+1)}{K}]}{\sqrt{K}} = \frac{\sqrt{2}\, \sin[\pi (r-\frac{\{Kr\}}{K} +\frac{1}{K})  (K x+1)]}{\sqrt{K}} ~.
\end{equation}
Consequently, using the definition of $G_K(x,R)$ in \eqref{GKFunction}, its leading form in the bulk region is
\begin{equation}
 G_K(x,r) = \frac{2}{K}\,\sin\left[\pi \left(r-\frac{\{Kr\}}{K} +\frac{1}{K} \right)  (K x+1)\right]\, \sin[\pi x] \,\exp[i K \left(\frac{n\pi  x}{K}-\frac{n\pi }{4 K}+\frac{n\pi  x^2}{2}\right)] ~.
 \label{GKBulk}
\end{equation}
Applying the Poisson summation formula, the function $F_R$ in \eqref{FuncFRSum}, along with \eqref{GDiscreteToCont}, will have the following asymptotics:
\begin{equation}
F_R = F_{\text{bulk}}(r)
\,\sim \, \left( 2^{\frac{1-d}{2}}\, \sin^{1-d}[\pi r]\, K^{\frac{(d-1)}{2}} \right) \left( K\,\sum_{u\,\in\,\mathbb Z}\, \int_{0}^1 G_K(x,r)\, e^{-2\pi i K u x}\, dx \right) ~.
\end{equation}
Now, the argument of the first sine term in the function $G_K(x,r)$ in \eqref{GKBulk} is at $O(K)$ because 
\[ \left(r-\frac{\{Kr\}}{K} +\frac{1}{K} \right)  (K x+1) = K r x + r +x -x \{Kr\} -\frac{\{Kr\}}{K}+\frac{1}{K} ~. \]
This makes this sine term highly oscillatory in the large-$K$ limit, and it must be included in the rapidly varying complex exponential. For this, we write:
\begin{equation}
\sin\left[\pi \left(r-\frac{\{Kr\}}{K} +\frac{1}{K} \right)  (K x+1)\right] = \frac{e^{i\pi \left( K r x-\frac{\{Kr\}}{K}+\frac{1}{K}+r-x \{Kr\}+x \right) } - e^{-i\pi \left( K r x-\frac{\{Kr\}}{K}+\frac{1}{K}+r-x \{Kr\}+x \right) }}{2i} \nonumber  ~.
\end{equation}
Substituting this decomposition into the Poisson-resummed expression gives two contributions:
\begin{equation}
F_{\text{bulk}}(r)
\,\sim \, \frac{ \sin^{1-d}[\pi r] \, K^{\frac{(d-1)}{2}}}{2^{\frac{d-1}{2}}\,i} \,\, \left(  \sum_{u\,\in\,\mathbb Z}I^{(1)}_u -  \sum_{u\,\in\,\mathbb Z} I^{(2)}_u \right) ~,
\label{Fbulkr}
\end{equation}
where the two integrals are given as:
\begin{equation}
I^{(1)}_u = \int_{0}^1 \sin[\pi x]\, e^{iK\,\Phi_{+}(x)}\, dx \quad;\quad I^{(2)}_u = \int_{0}^1 \sin[\pi x]\, e^{iK\,\Phi_{-}(x)}\, dx ~.
\end{equation}
The terms that are suppressed by $1/K$ in the phase do not affect the leading-order stationary points. Hence the two phase functions, after omitting such terms, will be given as:
\begin{align}
\Phi_{+}(x) = \frac{n \pi  x^2}{2} + \pi  r x - 2\pi u x \quad,\quad \Phi_{-}(x) = \frac{n \pi  x^2}{2} - \pi  r x - 2\pi u x ~.
\end{align}
For a fixed $r \in (0,1)$ and a given value of $n$, the stationary points for the two phase functions, determined by $\Phi_+'(x_+)=0$ and $\Phi_-'(x_-)=0$, are 
\[ x_{+} = \frac{2u-r}{n} \quad,\quad x_{-} = \frac{2u+r}{n} \]
Requiring these stationary points to lie in the interior of the integration interval $0<x_\pm<1$ restricts the allowed integer values of $u$ in the two sums in \eqref{Fbulkr}. Since both phase functions satisfy
\[
\Phi_\pm''(x)=n\pi ~,
\]
each interior stationary point gives a contribution of order $1/K^{1/2}$, which follows from the  stationary-phase formula \eqref{Istationaryphase}. In particular, the leading contributions are
\begin{align}
I^{(1)}_u & \sim \frac{\sqrt{2}\,e^{iK \Phi_+(x_+)}}{\sqrt{K \Phi_+''(x_+)}}\,e^{\frac{i \pi}{4}}\sqrt{\pi} \, \sin\left[\pi x_+\right] = \frac{\sqrt{2}\, }{\sqrt{Kn}}\,e^{-\frac{i \pi  K (r-2 u)^2}{2 n}}\,e^{\frac{i \pi}{4}} \, \sin\left[\frac{\pi(2u-r)}{n}\right] = \frac{C_+(u,r,n)}{\sqrt{K}} \nonumber \\ 
I^{(2)}_u & \sim \frac{\sqrt{2}\,e^{iK \Phi_-(x_-)}}{\sqrt{K \Phi_-''(x_-)}}\,e^{\frac{i \pi}{4}}\sqrt{\pi} \, \sin\left[\pi x_-\right] = \frac{\sqrt{2}\, }{\sqrt{Kn}}\,e^{-\frac{i \pi  K (r+2 u)^2}{2 n}}\,e^{\frac{i \pi}{4}} \, \sin\left[\frac{\pi(2u+r)}{n}\right] = \frac{C_-(u,r,n)}{\sqrt{K}} ~.
\label{Intu1and2}
\end{align}
Here $C_{\pm}$ denote the corresponding stationary-phase coefficients. These coefficients depend only on $K$ modulo $n$ via the phase term, but their magnitudes are independent of $K$. Hence, the two integrals asymptotically behave as $1/K^{1/2}$. Substituting these in \eqref{Fbulkr}, we get
\begin{equation}
F_{\text{bulk}}(r)
\,\sim \, \frac{ \sin^{1-d}[\pi r] \, K^{\frac{(d-2)}{2}}}{2^{\frac{d-1}{2}}\,i} \,\, \left[  \sum_{u} C_+(u,r,n) -  \sum_{u} C_-(u,r,n) \right] ~,
\label{}
\end{equation}
where the two sums run over appropriate integers $u$ corresponding to their interior stationary points, as we have discussed above. Hence, for fixed $r\in(0,1)$, the leading bulk asymptotics can be expressed as
\begin{equation}
F_{\text{bulk}}(r)
\,\sim \, C(r,n)\,K^{\frac{d-2}{2}} ~,
\end{equation}
where
\begin{equation}
C(r,n)
= \frac{ \sin^{1-d}[\pi r] }{2^{\frac{d-1}{2}}\,i} \,\, \left[  \sum_{u} C_+(u,r,n) -  \sum_{u} C_-(u,r,n) \right] ~.
\label{CoeffCr}
\end{equation} 
Since $\Lambda_R = |F_R|^2$, the bulk eigenvalues at fixed $r$ therefore have the leading scaling
\begin{equation}
\Lambda_R \equiv \Lambda_{\text{bulk}}(r) \sim |C(r,n)|^2 \,K^{d-2} ~.
\end{equation}
By contrast, the left-edge and right-edge eigenvalues scale as
\[
\Lambda_{\mathrm{left}}\sim K^{3d-4},
\qquad
\Lambda_{\mathrm{right}}\sim K^{3d-4},
\]
as obtained in \eqref{LeftEigenAsymp} and \eqref{RightEigenAsymp}, respectively. Since $(3d-4)>(d-2)$ for all $d \geq 2$, the bulk eigenvalues grow more slowly with $K$ than the edge eigenvalues. Since $\Lambda_R =O(K^{d-2})$ for bulk representations $R$, the number of such representations is at most $O(K)$, the bulk contribution to the trace satisfies
\begin{equation}
\text{Tr}[\sigma^m]_{\text{bulk}} = \sum_{R \, \in \, \text{bulk}} (\Lambda_{R})^m = O(K)\,O(K^{m(d-2)}) = O(K^{dm-2m+1}) ~.
\end{equation}
On the other hand, the left- and right-edge contributions scale as
\[
\text{Tr}[\sigma^m]_{\text{left}} \sim K^{3dm-4m} \quad,\quad
\text{Tr}[\sigma^m]_{\text{right}} \sim K^{3dm-4m} ~.
\]
Since $3dm-4m > dm-2m+1$ for all $d\geq 2$ and $m \geq 1$, the bulk contribution is subleading compared with the edge contributions.  
\subsection{Leading asymptotics and residue-class dependence}
Having determined the leading large-$K$ asymptotics of the $\text{Tr}[\sigma^m]$ in the left-edge, right-edge, and bulk regions, we can combine these contributions to obtain the leading asymptotic behavior of the moments. Since the bulk contribution is subleading, we have
\begin{equation}
\text{Tr}[\sigma^m] \,\, \sim \,\, \text{Tr}[\sigma^m]_{\text{left}} + \text{Tr}[\sigma^m]_{\text{right}}  ~.
\end{equation}
Adding the two edge contributions obtained in \eqref{TrSigmaLeftFinal} and \eqref{TrSigmaRightFinal}, we find
\begin{equation}
\text{Tr}[\sigma^m] \,\, \sim \,\, \frac{C_m}{2^{dm-4m}\,\pi^{2dm-2m}\,n^{2dm-m}} \,K^{3dm-4m}   ~,
\end{equation}
where
\begin{equation}
C_m =  \sum_{q=0}^{n-1} \left[ X(q)^m + Y(q)^m\right] \, \zeta\left(2dm-2m,\,\frac{q+1}{n}\right)  ~.
\label{CoeffCmfinal}
\end{equation}
Here $X(q)$ and $Y(q)$ are defined in \eqref{FuncXq} and \eqref{FuncYq} respectively. These functions contain $K$-dependent oscillatory phase factors $\exp[-2 i \pi  K s^2/n]$ and $\exp[-i \pi  K (2t-1)^2/n]$ respectively. Since $s$ and $t$ are integers, their values depend on the congruence class of $K$ modulo $n$. Since our primary interest is in the semiclassical limit of the entropies as the Chern–Simons level $k$ tends to infinity, we express this dependence in terms of the residue class of $k$. To this end, we write
\[
k=nk'+\ell,\qquad \ell=0,1,\ldots,n-1 ~,
\]
and take $k' \to \infty$, with $\ell$ fixed. Along each such subsequence, $K=k+2$ has a fixed residue class modulo $n$, determined by $(\ell+2) \bmod n$. The coefficients $C_m$ therefore depend on $\ell$, and we denote them by $C_m^{(\ell)}$. Thus, although the leading power of $k$ is the same for all residue classes, its coefficient can depend on $\ell$. So we have
\begin{equation}
\boxed{ \text{Tr}[\sigma^m] \,\, \sim \,\, \frac{C_{m}^{(\ell)}}{2^{dm-4m}\,\pi^{2dm-2m}\,n^{2dm-m}} \,k^{3dm-4m} }   ~,
\label{MomentLargekAsymp}
\end{equation}
where
\begin{equation}
C_m^{(\ell)} = \sum_{q=0}^{n-1} \left[ X_\ell(q)^m+Y_\ell(q)^m\right] \, \zeta\left(2dm-2m,\,\frac{q+1}{n}\right)  ~,
\label{CoeffCmMoment}
\end{equation}
with $\ell$ dependent functions $X$ and $Y$ given as
\begin{align}
X_\ell(q) &= \left| \sum_{s=1}^{\left\lfloor \frac{n-1}{2}\right\rfloor}\, e^{-\frac{2 i \pi  (\ell+2) s^2}{n}} \sin\left[\frac{2 \pi  s}{n}\right] \sin\left[\frac{2 \pi  (q +1) s}{n}\right]  \right|^2 \nonumber \\
Y_\ell(q) &= \left| \sum_{t=1}^{\left\lfloor \frac{n}{2}\right\rfloor}\, e^{-\frac{2 i \pi  (\ell+2) t(t-1)}{n}} \sin\left[\frac{ \pi(2t-1)}{n}\right] \sin\left[\frac{ \pi  (q +1) (2t-1)}{n}\right]  \right|^2 ~.
\label{FuncXandY}
\end{align}
\section{Large-$k$ limit of the entanglement entropy for the $T_{d,\,dn}$ torus links} \label{sec4}
In this section, we use the asymptotic expressions for the density-matrix moments derived in the previous section to obtain the large-\(k\) limits of the Rényi and entanglement entropies for the \(T_{d,dn}\) torus-link states. Although the leading moments exhibit a common power-law dependence on \(k\), their coefficients retain a dependence on the residue class of \(k\) modulo \(n\), which can lead to distinct limiting values of the entropies along different residue-class subsequences. 

Using the leading large-$k$ asymptotics derived in \eqref{MomentLargekAsymp}, we see that
\begin{equation}
\frac{\text{Tr}[\sigma^m]}{\text{Tr}[\sigma]^m} \, \sim \,  \frac{C_m^{(\ell)}\,k^{3dm-4m}+ O(k^{3dm-4m-1})}{\left[C_1^{(\ell)}\,k^{3d-4} + O(k^{3d-5}) \right]^m} \, \xrightarrow[k\equiv\ell\,(\text{mod } n)]{k\to\infty} \, \frac{C_m^{(\ell)}}{\left[ C_1^{(\ell)} \right]^m} ~.
\end{equation}
The powers of $k$ cancel in this ratio. Consequently, the Rényi entropy has a finite semiclassical limit along each fixed residue-class subsequence:
\begin{equation}
\mathcal{R}_{m,\infty}^{(\ell)} \equiv \lim_{k \to \infty} \mathcal{R}_m^{(\ell)} = \lim_{k \to \infty} \frac{1}{1-m}\, \ln\left[ \frac{\text{Tr}[\sigma^m]}{\text{Tr}[\sigma]^m} \right] = \frac{1}{1-m}\, \ln \left[ C_m^{(\ell)}/\left( C_1^{(\ell)} \right)^m \right] ~.
\end{equation}
The entanglement entropy is obtained from the Rényi entropy by taking the limit $m\to 1$, which gives
\begin{equation}
    \text{EE}
    =-\left.\frac{d}{dm}\left[(1-m)R_m\right]\right|_{m=1}.
\end{equation}
Consequently, the large-$k$ limit of the entanglement entropy for the $T_{d,dn}$ torus-link states is
\begin{equation}
    \boxed{ \text{EE}_{\infty}^{(\ell)}  
    =\ln C_1^{(\ell)} - \frac{1}{C_1^{(\ell)}} \left. \frac{dC_m^{(\ell)}}{dm}
    \right|_{m=1} } ~,
    \label{EElargekgeneral}
\end{equation}
where $\ell = 0,1,2,\ldots,n-1$ labels the residue class $k\equiv \ell\pmod n$. The notation $\text{EE}_{\infty}^{(\ell)}$ allows for the possibility that the limiting entanglement entropy depends on this residue class. Although $C_m^{(\ell)}$ was initially defined for integer Rényi index $m$, its explicit expression involves powers $X_\ell(q)^m$ and $Y_\ell(q)^m$, together with a Hurwitz zeta function whose first argument depends continuously on \(m\). This expression therefore admits a natural continuation to real $m$ in a neighbourhood of $m=1$, allowing the derivative in \eqref{EElargekgeneral} to be evaluated.

The expressions derived above determine the large-\(k\) limits of the Rényi and entanglement entropies in terms of finite sums involving the coefficients \(X_\ell(q)\), \(Y_\ell(q)\), and the Hurwitz zeta function. In particular, they reveal how the limiting entropies depend on the link parameters and on the residue class of \(k\) modulo \(n\). In the following section, we evaluate these expressions for representative values of \(n\) and \(d\), and investigate the resulting patterns of residue-class degeneracies and the dependence of the limiting entanglement entropy on the number of link components.
\section{Examples and numerical results}\label{sec5}
In this section, we examine the analytical results obtained earlier through explicit evaluations and numerical comparisons. We first study the dependence of the limiting entanglement entropy on the residue class of \(k\) modulo \(n\), identifying cases in which different residue classes yield identical limiting values. We then investigate how these limiting values vary with the number of link components \(d\) and the winding parameter \(n\). These examples illustrate the structure of the large-\(k\) limits and provide a basis for examining the patterns that emerge as the link parameters are varied.
\subsection{Residue-class dependence and degeneracies}
\begin{table}[htbp]
    \centering
    \begin{tabular}{|c|c|c|} 
        \hline \rowcolor{Gray}
        $n$ & Distinct limiting values & Residue classes $\ell$ \\ \hline
        2            & 0.888            & $\{0,1\}$            \\ \hline
        3            & 1.938            & $\{0,1,2\}$            \\ \hline
        4 & 1.581,\, 1.524 & $\{0,2\}, \{1,3\}$ \\ \hline
        5 & 1.828,\, 2.439,\, 1.341 & $\{0,1\}, \{2,4\}, \{3\}$ \\ \hline
        6 & 1.388,\, 1.245,\, 1.929,\, 1.189 & $\{0,2\}, \{1\},\{3,5\},\{4\}$ \\ \hline
        7 & 1.774,\, 2.104,\, 2.777,\, 1.083 & $\{0,3\}, \{1,2\},\{4,6\},\{5\}$ \\ \hline
        8 & 1.581,\, 1.185,\, 0.888,\, 2.218,\, 1.007 & $\{0,4\}, \{1,3\},\{2\},\{5,7\}, \{6\}$ \\ \hline
        9 & 1.741,\, 1.938,\, 2.342,\, 3.030,\, 0.952  & $\{0,5\}, \{1,4\},\{2,3\},\{6,8\}, \{7\}$ \\ \hline
        10 & 1.466,\, 1.365,\, 1.231,\, 0.651,\, 2.443,\, 0.911  & $\{0,6\}, \{1,5\},\{2,4\}, \{3\}, \{7,9\}, \{8\}$ \\ \hline
          \end{tabular}
    \caption[Values of limiting entanglement entropy for various residue classes for a given $n$.]{Distinct limiting values of entanglement entropy for different values of $n$, together with the corresponding degenerate residue classes. We have fixed $d=2$. The limiting values of EE are calculated using Eq. \eqref{EElargekgeneral} and reported to three decimal places.}
    \label{tab:grid_table}
\end{table}
In Table~\ref{tab:grid_table}, we present the limiting entanglement entropies $\text{EE}_{\infty}^{(\ell)}$, calculated using \eqref{EElargekgeneral}, for various values of $n$ and $\ell$, with $d=2$. For a given $n$, different residue classes $\ell$ can yield identical limiting entropy values. We group such residue classes together and list only the distinct limiting values. For example, for $n=5$, we obtain 
\begin{equation}
\mathrm{EE}_{\infty}^{(\ell)}(n=5,\,d=2) \, \approx \, \begin{cases}
    1.828 & \text{,\,\, for $\ell=0, 1$} \\ 2.439 & \text{,\,\, for $\ell=2, 4$} \\ 1.341 & \text{,\,\, for $\ell=3$}
\end{cases}  ~.
\end{equation}
\begin{figure}[htb]
    \centering
    \includegraphics[width=1.0\textwidth]{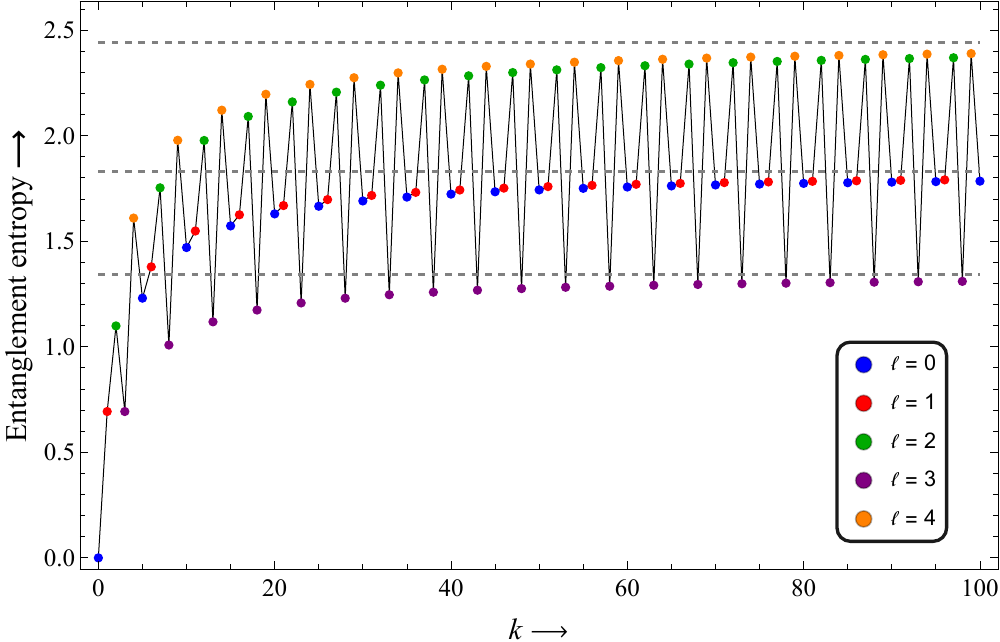}
    \caption[Entanglement entropy as a function of the level $k$ for $T_{2,10}$ link.]{Entanglement entropy as a function of the level $k$ for $n=5$ and $d=2$. The colored markers distinguish the residue classes corresponding to $k = \ell  \bmod 5$. The horizontal dashed lines indicate the three distinct limiting values, $\mathrm{EE}_{\infty}^{(\ell)}\approx 1.828$, $2.439$, and $1.341$, corresponding to the residue-class groups $\{0,1 \}$, $\{2,4\}$, and $\{3\}$, respectively.}
    \label{PlotEEn5d2}
\end{figure}
Figure~\ref{PlotEEn5d2} shows the variation of the entanglement entropy with the level $k$ for $n=5$ and $d=2$. The black curve connects the entropy values obtained at successive levels, while the coloured markers distinguish the five residue classes $k=\ell \bmod 5$. As $k$ increases, the entropy exhibits a clear convergence pattern: the five residue-class subsequences approach three distinct limiting values, approximately $1.828$, $2.439$, and $1.341$. Specifically, the classes $\ell=0,1$ converge to $1.828$, the classes $\ell=2,4$ converge to $2.439$, and the class $\ell=3$ converges to $1.341$, as indicated by the horizontal dashed lines. This behavior illustrates how the residue-class dependence persists in the large-$k$ limit, while distinct residue classes can yield identical limiting entanglement entropies.

\subsection{Variation of limiting entanglement entropy with winding parameter}
\begin{figure}[htb]
    \centering
    \includegraphics[width=1.0\textwidth]{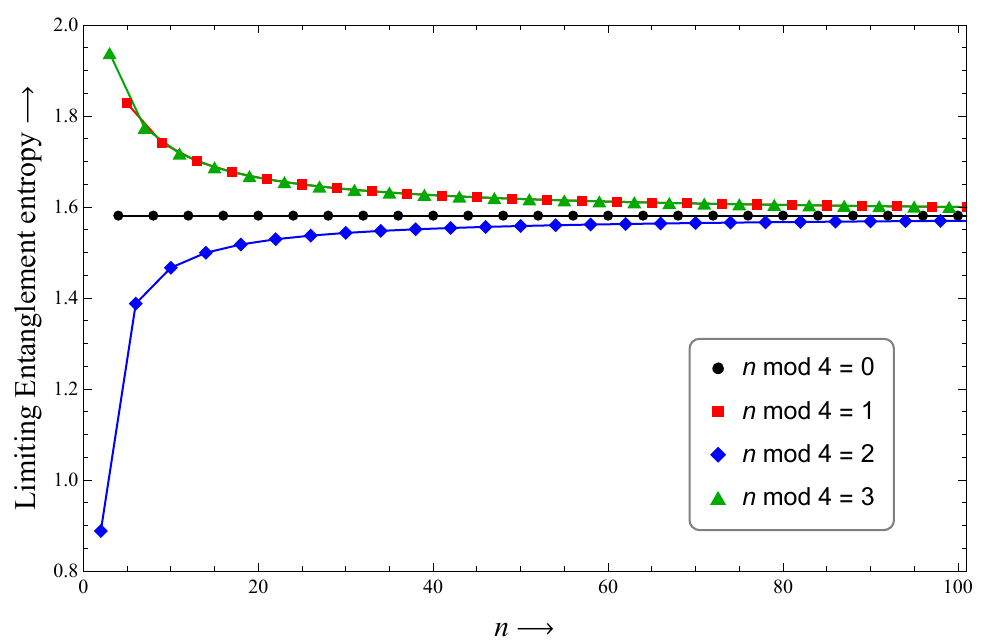}
    \caption[Limiting entanglement entropy as function of winding parameter $n$ for $T_{2,\,2n}$ link.]{Limiting entanglement entropy $\mathrm{EE}_{\infty}^{(0)}$ as a function of the winding parameter $n$ for $d=2$. The data are grouped according to the residue class $n\bmod 4$. When $n$ is a multiple of 4, the limiting entropy is independent of $n$ and approximately equals $1.58099$. The other three subsequences approach this value as $n$ increases, as suggested by the numerical results.}
    \label{PlotEElimitFORd2l0}
\end{figure}

In this subsection, we investigate the dependence of the limiting entanglement entropy $\mathrm{EE}_{\infty}^{(\ell)}$ on the winding parameter $n$, keeping the number of link components fixed at $d=2$. We focus on the residue class $\ell=0$, which corresponds to restricting the level $k$ to integer multiples of $n$ and then taking the limit $k \to \infty$. Using the analytical expression derived in \eqref{EElargekgeneral}, we evaluate $\mathrm{EE}_{\infty}^{(0)}$ for $n=2,3,\ldots,100$, and the values are presented in the  Figure~\ref{PlotEElimitFORd2l0}.

Figure~\ref{PlotEElimitFORd2l0} exhibits four distinct patterns corresponding to the subsequences classified by $n\bmod 4$. For $n=4\mathbb{Z}$, the limiting entanglement entropy is independent of $n$ and is given exactly by
\[
\mathrm{EE}_{\infty}^{(0)} = -2\gamma-\frac{14}{3}\ln 2
+24\ln A ~,
\]
where $\gamma$ is the Euler--Mascheroni constant and $A$ is the Glaisher--Kinkelin constant. Numerically, this gives $\mathrm{EE}_{\infty}^{(0)}\approx 1.58099$. The subsequences with $n=4\mathbb{Z}+1$ and $n=4\mathbb{Z}+3$ decrease monotonically towards this value as $n$ increases, whereas the subsequence with $n=4\mathbb{Z}+2$  increases from below and gradually approaches it. Although the four subsequences differ appreciably at small $n$, their values become progressively closer as $n$ increases. The numerical results up to $n=100$ therefore suggest that all four subsequences converge to the same value. This behavior indicates that the dependence on the residue class of $n$, which is prominent at small $n$, becomes less pronounced in the large-$n$ regime. 

\subsection{Variation of limiting entanglement entropy with number of link components}
\begin{figure}[htb]
    \centering
    \includegraphics[width=1.0\textwidth]{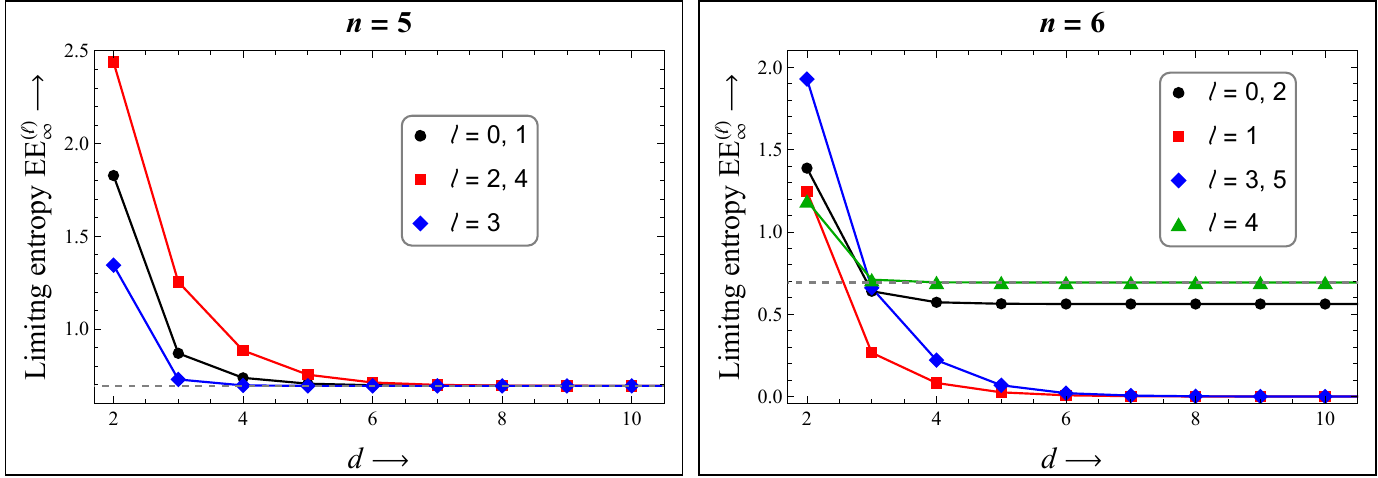}
    \caption[Limiting entanglement entropy as a function of $d$ for $T_{d,\,5d}$ and $T_{d,\,6d}$ links.]{Limiting entanglement entropy $\mathrm{EE}_{\infty}^{(\ell)}$
as a function of the number of link components $d$ for
$n=5$ (left panel) and $n=6$ (right panel). Different curves
correspond to the residue classes $\ell\equiv k\, (\bmod n)$. The horizontal dashed line indicates the value $\ln 2$.}
    \label{PlotEElimitFORn5n6}
\end{figure}
In this subsection, we investigate the dependence of the limiting
entanglement entropy $\mathrm{EE}_{\infty}^{(\ell)}$ on the number
of link components $d$, for fixed winding parameter $n$ and residue
class $\ell$ of the level $k$ modulo $n$. We consider the cases
$n=5$ and $n=6$, for which we evaluate
$\mathrm{EE}_{\infty}^{(\ell)}$ for $d=2,\ldots,10$ using the
analytical expression derived in \eqref{EElargekgeneral}.
The resulting behavior is shown in Figure~\ref{PlotEElimitFORn5n6}.
The numerical results reveal a clear dependence on the parity of $n$ and on the residue class $\ell$, and suggest a connection with the large-party limit studied in \cite{Sain:2026ryl}. We will explore these next.

For $n=5$, the limiting entanglement entropy $\mathrm{EE}_{\infty}^{(\ell)}$ for all residue classes
$\ell=0,1,2,3,4$ approaches $\ln 2$ as $d$ increases. In contrast,
for $n=6$, the subsequences corresponding to
$\ell=1,3,5$ approach zero, while the remaining residue classes
exhibit a different behaviour. These observations indicate that
the large-$d$ limit depends on both the winding parameter and the
residue class of the level. In particular, the numerical results
suggest that the limiting entropy approaches $\ln 2$ for odd $n$, whereas for even $n$
it vanishes for odd residue classes $\ell$. To understand these
patterns, we next derive the large-$d$ limit
directly from the limiting density-matrix moments.

\paragraph{Analytical large-$d$ limit.}
To derive the large-$d$ behavior, we start from the leading order coefficient in the large-$k$ expansion of the moments $\text{Tr}[\sigma^m]$. These coefficients $C_m^{(\ell)}$ are given in \eqref{CoeffCmMoment}. The $d$-dependence in these coefficients comes explicitly from the Hurwitz zeta function, 
$$ \zeta\left(
2m(d-1),\frac{q+1}{n}
\right) ~. $$
Now, for $\zeta(s,a)$ with $s \to \infty$ and a fixed $a>0$, the leading order behavior is given as
\begin{equation}
\zeta(s,a)=\sum_{i=0}^{\infty}\frac{1}{(a+i)^s} \, \xrightarrow{s \, \to \, \infty} \, \frac{1}{a^s} ~.
\end{equation}
Since $m \geq 1$, the first argument of the Hurwitz zeta function $s=2m(d-1) \to \infty$ as $d\to \infty$. Thus, for a given value of $q$ and $n$, the large-$d$ expansion can be
\begin{equation}
\zeta\left(
2m(d-1),\frac{q+1}{n}
\right)
\sim
\left(\frac{n}{q+1}\right)^{2m(d-1)} ~.
\label{zeta_asymptotic_d}
\end{equation}
Hence, from \eqref{CoeffCmMoment}, the large-$d$ behavior of $C_m^{(\ell)}$ will be dominated by the $q=0$ term, provided that $X_\ell(0) + Y_\ell(0) > 0$, which we have verified for various $n$ and $\ell$ values. The subleading contributions from $q\geq1$ are exponentially suppressed relative to the $q=0$ contribution as $d\to\infty$, and remain negligible upon differentiation with respect to $m$ in a neighborhood of $m=1$. Thus, in the large-$d$ regime, we will have:
\begin{equation}
C_m^{(\ell)}
\sim
n^{2m(d-1)}
\left(X_\ell(0)^m+Y_\ell(0)^m\right) ~,
\label{Cm_d_asymptotic}
\end{equation}
where, 
\begin{equation}
X_\ell(0) = \left| \sum_{s=1}^{\left\lfloor \frac{n-1}{2}\right\rfloor}\, e^{-\frac{2 i \pi  (\ell+2) s^2}{n}} \sin^2\left[\frac{2 \pi  s}{n}\right]  \right|^2 \quad;\quad
Y_\ell(0) = \left| \sum_{t=1}^{\left\lfloor \frac{n}{2}\right\rfloor}\, e^{-\frac{2 i \pi  (\ell+2) t(t-1)}{n}} \sin^2\left[\frac{ \pi(2t-1)}{n}\right]  \right|^2 ~.
\label{FuncXandYq0}
\end{equation}
Substituting this into the \eqref{EElargekgeneral}, the large-$d$ limit will be:
\begin{equation}
 \lim_{d\to\infty} \mathrm{EE}_{\infty}^{(\ell)}  
    =\ln\left[X_\ell(0) + Y_\ell(0) \right] - \frac{X_\ell(0) \ln X_\ell(0) + Y_\ell(0) \ln Y_\ell(0)}{X_\ell(0) + Y_\ell(0)}  ~.
    \label{}
\end{equation}
Introducing the normalized weights,
\begin{equation}
p_\ell=\frac{X_\ell(0)}{X_\ell(0)+Y_\ell(0)} \quad,\quad q_\ell= \frac{Y_\ell(0)}{X_\ell(0)+Y_\ell(0)} ~,
\end{equation}
with $p_{\ell}+q_{\ell}=1$, this limit can be written as:
\begin{equation}
\boxed{
\lim_{d\to\infty}\mathrm{EE}_{\infty}^{(\ell)}
= -(p_\ell\,\ln p_\ell + q_\ell\,\ln q_\ell ) } ~.
\label{EELimitLargekLarged}
\end{equation}
Thus, the large-$d$ limit of the limiting entanglement entropy is precisely the binary entropy associated with the normalized weights $p_\ell$ and $q_\ell$. Consequently, it satisfies
\[ 0 \leq \lim_{d\to\infty}\mathrm{EE}_{\infty}^{(\ell)} \leq \ln 2 ~.\]
The upper bound is attained when the two weights are equal, while the entropy vanishes when either of the two weights is zero.

\paragraph{Dependence on the parity of $n$.} When $n$ is odd, we have $X_\ell(0) = Y_\ell(0)$, thus the two weights $p_\ell$ and $q_\ell$ are equal. Hence, the large-$d$ limit of the limiting entanglement entropy achieves its maximum value of $\ln 2$. This is also evident from the $n=5$ plot in Figure~\ref{PlotEElimitFORn5n6}. To prove that $X_\ell(0)$ and $Y_\ell(0)$ are identical for all odd $n$ values, we write the two coefficients in \eqref{FuncXandYq0} as:
\begin{equation}
X_\ell^{\text{odd}}(0) = \left| \sum_{s=1}^{\frac{n-1}{2}}\, e^{-\frac{2 i \pi  (\ell+2) s^2}{n}} \sin^2\left[\frac{2 \pi  s}{n}\right]  \right|^2 \quad;\quad
Y_\ell^{\text{odd}}(0) = \left| \sum_{t=1}^{ \frac{n-1}{2} }\, e^{-\frac{2 i \pi  (\ell+2) t(t-1)}{n}} \sin^2\left[\frac{ \pi(2t-1)}{n}\right]  \right|^2 ~.
\label{FuncXandYq0nOdd}
\end{equation}
In $Y_\ell^{\text{odd}}(0)$, we make the change of summation
variable $s=\frac{n+1}{2}-t$. Under this transformation, $2t-1 = n-2s$ and $t(t-1) = s^2-ns+\frac{n^2-1}{4}$. It follows that the sine-squared factor becomes identical to
that in $X_\ell^{\text{odd}}(0)$, while the exponential factor
differs only by an overall phase independent of $s$.
Since this phase has unit modulus, we obtain $X_\ell^{\text{odd}}(0)=Y_\ell^{\text{odd}}(0) > 0$. As a result, $p_\ell = q_\ell = 1/2$ and we get
\[ \lim_{d\to\infty} \mathrm{EE}_{\infty}^{(\ell)} = \ln 2 \,, \qquad (\text{for odd } n) ~. \]

On the other hand, when $n$ is even and $k \equiv \ell = \text{odd } (\bmod n)$, one of the coefficients $X_\ell(0)$ or $Y_\ell(0)$ vanishes, and hence the corresponding large-$d$ limit of the limiting entanglement entropy vanishes. This is clear from the $n=6$ plot in Figure~\ref{PlotEElimitFORn5n6}. To see this behavior, consider $n=2u=\text{even}$ and write the two coefficients as $X_\ell^{\text{even}}(0) = |S_X|^2$ and $Y_\ell^{\text{even}}(0) = |S_Y|^2$, where:
\begin{equation}
S_X = \sum_{s=1}^{u-1}\, e^{-\frac{i \pi  (\ell+2) s^2}{u}} \sin^2\left[\frac{\pi  s}{u}\right] \quad;\quad
S_Y = \sum_{t=1}^{u}\, e^{-\frac{i \pi  (\ell+2) t(t-1)}{u}} \sin^2\left[\frac{ \pi(2t-1)}{2u}\right] ~.
\label{FuncXandYq0nEven}
\end{equation}
In $S_X$, we make the change of summation variable $s'=u-s$. Using the invariance of the sine-squared factor under this transformation, we obtain $S_X = e^{-i\pi \ell u}S_X$. Similarly, in $S_Y$, we make the change of variable $t'=u+1-t$, which gives $S_Y = e^{-i\pi \ell (u+1)}S_Y$. We see that for odd $\ell$, $S_X$ vanishes for odd $u$ and $S_Y$ vanishes for even $u$. Consequently, for even $n$ and odd $\ell$, exactly one of $X_\ell(0)$ and $Y_\ell(0)$ vanishes, while the other is nonzero. We therefore obtain
\begin{equation}
\lim_{d\to\infty}\mathrm{EE}_{\infty}^{(\ell)}=0\,,
\qquad (\text{for } n\ \text{even and } \ell\ \text{odd}).
\end{equation}

\paragraph{The double limit commutes.} We now compare our results with those obtained in earlier work \cite{Sain:2026ryl}, where the large-party (large-$d$) limit of the entanglement entropy for torus links was studied first, followed by its semiclassical (large-$k$) limit for the class of $T_{d,\,dn}$ torus links. We find that our result in \eqref{EELimitLargekLarged} exactly reproduces the result obtained in Ref.~ \cite{Sain:2026ryl}. In particular, the two results take the form of a binary entropy involving two normalized weights, which precisely coincide with our weights $p_\ell$ and $q_\ell$. This agreement establishes that the two iterated limits commute: 
\begin{equation}
\boxed{ \lim_{d\to\infty} \left( \lim_{k\to\infty}\mathrm{EE} \right) = \lim_{k\to\infty} \left( \lim_{d\to\infty}\mathrm{EE} \right) = -(p_\ell\,\ln p_\ell + q_\ell\,\ln q_\ell ) } ~.
\label{}
\end{equation}

\section{Conclusion and Discussion} \label{sec6}
In this work, we have investigated the semiclassical asymptotics of entanglement entropy for quantum states associated with $T_{d,\,dn}$ torus links in SU(2)$_k$ Chern--Simons theory. Our analysis focused on the large-level limit $k\to\infty$, with the number of link components $d$ and the winding parameter $n$ held fixed. By exploiting the modular-matrix representation of the link-state coefficients, we derived the asymptotic behavior of the moments of the reduced density matrix and, consequently, obtained an analytical expression for the limiting entanglement entropy.

To determine the large-$k$ behavior of the moments, we divide the representation space into three regions: the left-edge region, where the representation label $R$ remains finite as $k\to\infty$; the right-edge region, where $k-R$ remains finite; and the bulk region, where $R$ scales with $k$, such that $R/k$ approaches a finite value strictly between zero and one. A central result of our analysis is that the leading asymptotic behavior of the density-matrix moments is governed by contributions from the two edges of the integrable-representation space, whereas the bulk contribution is subleading. Using these asymptotic expressions for the moments, we obtain an analytical expression for the large-$k$ limit of the entanglement entropy for $T_{d,\,dn}$ torus links in terms of finite sums involving Hurwitz zeta functions. Although the entropy converges in the large-$k$ limit, its limiting value generally depends on the residue class $\ell=k\bmod n$. This dependence reflects the sensitivity of the semiclassical limit to the arithmetic properties of the Chern--Simons level.

We have further explored the dependence of the limiting entanglement entropy on the winding parameter $n$, keeping the number of link components $d$ fixed. Our numerical results reveal that different residue classes can lead to distinct limiting values, with certain classes exhibiting degeneracies. The degenerate classes, together with the corresponding numerical values of the limiting entanglement entropy, are listed in Table~\ref{tab:grid_table} for several small values of $n$.

We have also investigated the behavior of the limiting entropy in the large-$d$ regime, with $n$ held fixed. By employing the large-$d$ asymptotic form of the Hurwitz zeta functions, we derive an analytical expression for this limit. As shown in Eq.~\eqref{EELimitLargekLarged}, the resulting expression takes the form of a binary entropy involving two normalized weights associated with the left- and right-edge contributions. We find that these weights are equal for odd $n$, yielding the maximum binary entropy $\ln 2$. For even $n$ and even residue classes $\ell$, one of the weights vanishes, resulting in a zero large-$d$ limit of the entropy.

An interesting outcome is the agreement between the two possible orders of taking the large-$k$ and large-$d$ limits. We have explicitly verified that the expression obtained by first taking the semiclassical limit ($k\to \infty$) and subsequently the large-party limit ($d\to \infty$) coincides with the result obtained by reversing the order of these limits, as previously studied in Ref.~\cite{Sain:2026ryl}. Thus, for the class of $T_{d,dn}$ links considered here, the two iterated limits commute. This agreement provides a nontrivial consistency check of our asymptotic analysis and connects the finite-party semiclassical description with the previously obtained large-party result.

The present analysis also raises several interesting questions for future investigation. It would be worthwhile to extend the asymptotic framework to more general torus links \(T_{p,q}\), for which the structure of the link-state coefficients may lead to a richer pattern of contributions from different regions of the representation space. Another natural direction is to investigate whether the edge-dominance mechanism and the associated residue-class dependence persist for other compact gauge groups and more general classes of link states. Finally, a systematic analytical study of the large-\(n\) behavior of the limiting entanglement entropy could provide a deeper understanding of the numerical trends observed in this work. These questions may help clarify the extent to which the asymptotic structure found here is a special feature of $T_{d,\,dn}$ links or reflects a more general property of topological entanglement in Chern--Simons theory.

\vspace{0.5in}

\noindent \textbf{Acknowledgment}\\
This work is supported by the `ANRF (erstwhile SERB)', Government of India, through the Start-up Research Grant (SRG) [Grant No. SRG/2023/001023]. VG acknowledges the `Council of Scientific and Industrial Research (CSIR)' for its support through the research fellowship.

\bibliographystyle{JHEP}
\bibliography{Torusppn}

\end{document}